\documentclass{elsarticle}
\journal{Journal of Computational Physics}
\usepackage{lineno}
\usepackage{amsmath}
\usepackage{amssymb}
\usepackage{subcaption}
\usepackage{xspace}
\usepackage{color}

\newcommand{\pT}{$p_{T}$\xspace}
\newcommand{\pTs}{$p_{T}$s\xspace}
\newcommand{\antikt}{anti-$k_{t}$\xspace}
\newcommand{\Znunu}{$Z\rightarrow\nu\bar{\nu}$\xspace}
\newcommand{\ETMiss}{$E_{T}^{\mathrm{miss}}$\xspace}
\newcommand{\avnp}{$\left<\mu\right>$\xspace}
\newcommand{\rms}[2]{C^{\mathrm{RMS}}_{#1,#2}\xspace}
\newcommand{\avrmsmu}[2]{$\langle C^{\mathrm{RMS}}_{#1,#2}\rangle/\sqrt{\mu}$\xspace}
\newcommand{\ly}{l_{y}}
\newcommand{\lphi}{l_{\phi}}

\DeclareMathOperator\erf{erf}
\begin{document}

\begin{frontmatter}
  
  \title{A Wavelet Based Pile-Up Mitigation Method for the LHC Upgrade}
  

  \author[add1]{P.~Hansen}
  \author[add1,add2]{J.~W.~Monk\corref{ead}}
  \cortext[ead]{primary author (jmonk@cern.ch)}
  \author[add1]{C.~Wiglesworth}
  \address[add1]{Niels Bohr Institute, Blegdamsvej 17, Copenhagen, Denmark}
  \address[add2]{Dept. of Astronomy and Theoretical Physics, S\"{o}lvegatan 14A, Lund, Sweden}

  \begin{abstract}
    Collision experiments at the Large Hadron Collider suffer from the problem of pile-up, which is the read-out of multiple simultaneous background proton-proton collisions per beam-crossing. We introduce a pile-up mitigation technique based on wavelet decomposition. Pile-up is treated as a form of white noise, which can be removed by filtering beam-crossing events in the wavelet domain. The particle-level performance of the method is evaluated using a sample of simulated proton-proton collision events that contain $Z$ bosons decaying to a pair of neutrinos, overlaid with pile-up. In the wavelet representation, the pile-up noise level is found to grow with the square root of the number of background proton-proton collisions.
  \end{abstract}
  
  
\end{frontmatter}

\setlength{\abovedisplayskip}{5pt}
\setlength{\abovedisplayshortskip}{5pt}

\setlength{\belowdisplayskip}{15pt}
\setlength{\belowdisplayshortskip}{15pt}


\section{Introduction}
\label{sec:introduction}

\subsection{Pile-Up at the Large Hadron Collider}

Run-2 of the Large Hadron Collider (LHC) has seen an increase in both the centre-of-mass energy and the instantaneous luminosity, with respect to Run-1. As a result, the number of proton-proton collisions per beam-crossing ($\mu$) has also increased. At the time of writing the mean value of $\mu$ over the Run-2 data taking period has been approximately 35 \cite{twiki:lumi}.

There are a number of planned upgrades for the LHC, from which will see a further increase in the centre-of-mass energy and the instantaneous luminosity. In particular, the High-Luminosity LHC (HL-LHC) is due to begin in 2026. It will operate at an instantaneous luminosity of up to 7.5 times the nominal design specification of the LHC and $\mu$ is expected to reach values of up to 200 \cite{ApollinariG.:2017ojx}. It aims to deliver a (total) integrated luminosity of up to 3000 fb$^{-1}$ and will permit a wide range of rare processes to be studied with precision measurements.  Some of the important physics processes that will be opened up by this large dataset include $H\rightarrow\mu\mu$, vector boson fusion (VBF) production of $H\rightarrow\gamma\gamma$ and $H\rightarrow\tau\tau$, and the Higgs self-coupling \cite{atlas:loi}. However, the high instantaneous luminosities that enable these measurements also make the observation of individual events much more difficult because the large number of overlaid multiple soft collisions either obscure some of the process-specific signatures, or degrade the observation efficiency and resolution.  The efficacy of the methods used to mitigate the effects of the multiple proton-proton collisions (henceforth pile-up) will therefore be one of the important limiting factors of the high luminosity physics programme.


A range of pile-up mitigation methods have already  been  employed during the first two runs of the LHC.  Some of these approaches are observable specific, such as jet area subtraction \cite{Cacciari:2007fd, Soyez:2012hv} or jet substructure techniques and trimming \cite{Butterworth:2008iy, Asquith:2018igt, Krohn:2009th}.  Other methods such as SoftKiller \cite{Cacciari:2014gra} or PUPPI \cite{Bertolini:2014bba} have sought to classify individual particles as originating from either signal or pile-up processes.  These particle-classifying methods define a unique angular scale that is expected to characterise the pile-up.  In the case of SoftKiller the scale enters as a grid size, while for PUPPI the scale is defined by the radius of a cone around each candidate particle.

In this paper, we introduce a new approach for identifying individual pile-up particles based on wavelets and implement two algorithms using that method.  Pile-up is modelled as a form of white noise whose angular scaling properties are exploited in order to remove it. This method is intrinsically multi-scale and does not require any characteristic angular selection for pile-up, while at the same time it makes use of information from the entire event, not just local regions close to target emissions.  The performance of the algorithms for the retention and rejection of individual signal and pile-up particles is evaluated on a sample of simulated proton-proton collision events that contain $Z$ bosons decaying to a pair of neutrinos. The neutrino sample is deliberately chosen as a challenge for pile-up removal, since the hard signal activity is not visible, thus any QCD activity in the signal collision has similar characteristics to soft QCD activity in pile-up. The scope of this paper is to provide a concise description and introduction to the method, as well as a brief demonstration that it works even in challenging pile-up scenarios.  We have therefore not included an evaluation on a wider range of processes, nor a comparison to other pile-up mitigation methods.  Such an in depth evaluation of the performance of the method will be presented in future work.

\subsection{Noise Suppression and Pile-Up Mitigation}\label{sec:noise}

A sample, $N(\phi)$, composed of white noise shows no correlation between the values of $N$ at two different values of $\phi$. When viewed in the frequency domain, such a noise sample will show a power spectrum that is flat as a function of frequency. 

With a sufficiently large number of overlaid proton-proton collisions, the particles observed during a single beam-crossing event at the LHC are uncorrelated with the majority of the other particles due to the fact that they originate from many independent collisions. To a good approximation, multiple proton-proton collisions can therefore be considered as a form of white noise, a concept that is explored in Section~\ref{sec:softqcd}. 

On the other hand, a \emph{single} high transverse momentum signal collision produces strongly correlated emissions of particles, which is expected to result in a power spectrum that is sparsely populated in the frequency domain. Since we expect that the signal can be sparsely encoded (i.e. compressed) in the frequency domain, and that the noise cannot, this forms the basis of a method for separating signal particles from pile-up.

\begin{figure}[t!]
  \centering
  \includegraphics[width=0.5\textwidth]{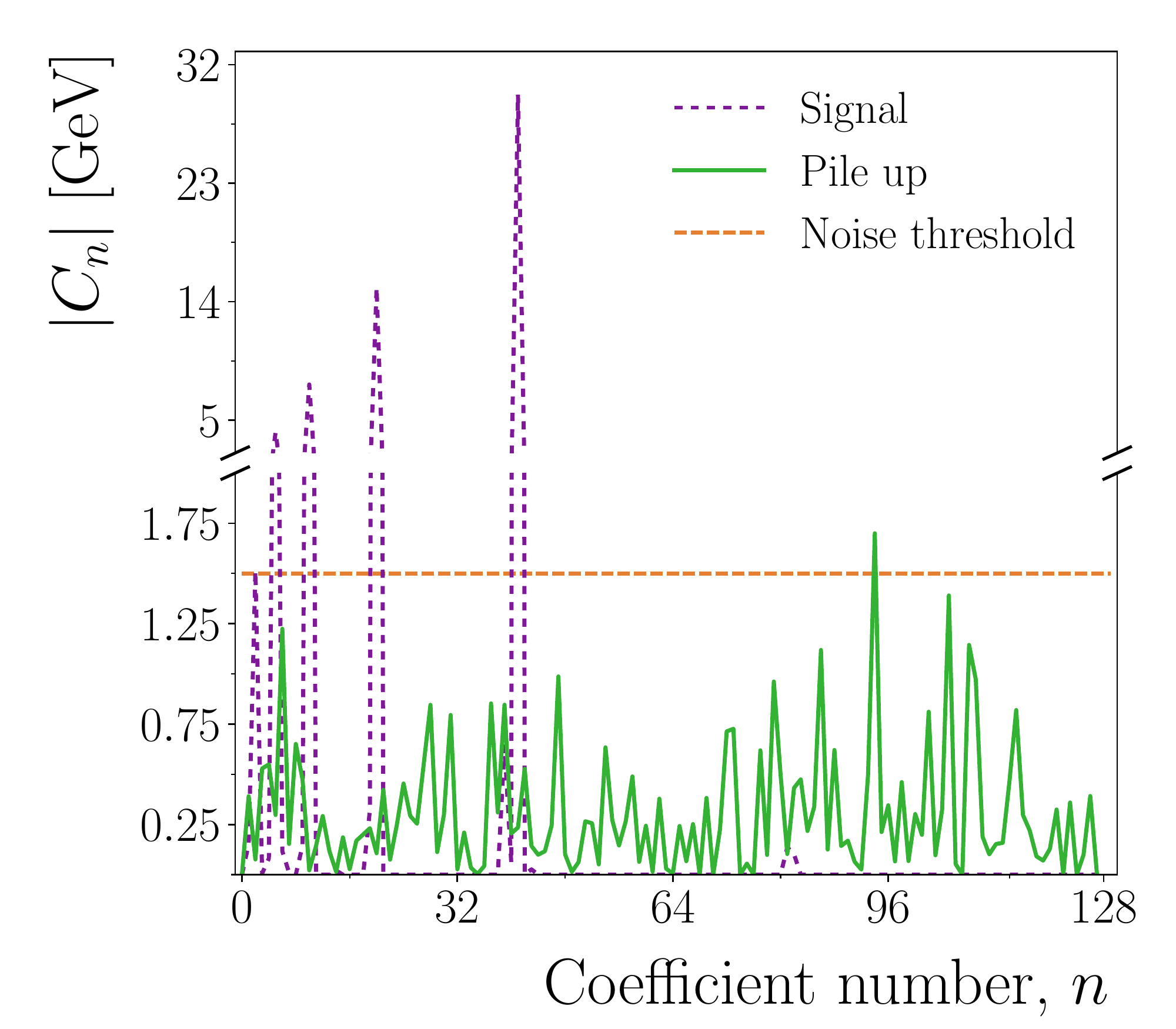}
  \caption{A representation of the frequency spectrum of signal and noise components of a QCD jet event at $\mu$=100. The noise component (green) fluctuates around a mean value, while the signal (dashed purple) populates a small number of spikes that extend above the noise level. The orange line is a suggested threshold for filtering the noise component.  Note the change of scale on the vertical axis, which is employed to show both the details of the noise and the extent of the signal spikes.}
  \label{fig:schematic}
\end{figure}

This is illustrated in Figure~\ref{fig:schematic}, which uses a single QCD production signal event overlaid with 100 soft QCD events, simulated using Pythia 8 \cite{Sjostrand:2006za, Sjostrand:2014zea}.  The signal event is selected to contain three \antikt jets with radius $R=0.4$ and transverse momentum \pT$ > 100$~GeV.  Figure \ref{fig:schematic} shows the result of transforming the event from rapidity-azimuth ($y - \phi$) to the wavelet domain\footnote{where we have previously used the frequency domain to refer to a general event transformation in abstract terms, we now use the term wavelet domain to refer to the concrete example of the wavelet transform.}, which is described by a set of coefficients, $C_{n}$.  The nature of the transformation is explained in Section \ref{sec:transform}, but it can be seen that the pile-up contribution populates the frequency space uniformly while the jet signal appears as a small number of large spikes that are an order of magnitude higher than the pile-up.

The observation that the signal is encoded in a small number of coefficients and that the noise is uniform leads to a simple noise removal prescription. A threshold, $T$, is specified, and any contribution that lies below that threshold in the wavelet domain is removed. An example threshold is shown as the orange line in Figure \ref{fig:schematic}.  This means that each coefficient, $C_{n}$, in the decomposition is replaced by a new value, $C_{n}^{\prime}$, given by equation \ref{eqn:cprime}

\begin{equation}
  C_{n}^{\prime} = 
  \begin{cases}
    C_{n}, & \text{if}\ \left|C_{n}\right| > T \\ 
    0,     & \text{otherwise}
  \end{cases}
  \label{eqn:cprime}
\end{equation}

If a threshold can be chosen that is above the noise level and sufficiently below the signal peak, the majority of the removed activity will be noise. Any signal that lies below the noise threshold is removed, and the small part of the noise that is coincident with the signal in the wavelet domain will be retained. All pile-up mitigation methods suffer from such overlap between signal and noise, and maximising the separation between the two is the key to a performant algorithm.

\subsection{Discrete Wavelet Transformation of Collision Events}\label{sec:transform}

There are a large range of transformations that could be applied to proton-proton collision events in order to reveal their characteristics in an alternative domain. For this new mitigation algorithm we will use the two-dimensional discrete wavelet transformation with a dyadic\footnote{In the dyadic hierarchy, the angular resolution halves for each level of decomposition.  Alternative decompositions are possible, but the dyadic representation is most commonly used and is easily accessible with software  tools.} decomposition. An advantage of using a wavelet transformation is that encoding the signal using a wavelet basis provides a sparse representation of the signal. This sparse representation results in less overlap between the noise and the signal in the wavelet domain and, consequently, less pile-up contamination after thresholding. In addition, the wavelet basis functions have compact support, which reduces the size of edge effects due to the finite pseudo-rapidity coverage of collider detectors.  

Wavelet analysis is a general method for analysing the scaling behaviour of a system with many degrees of freedom, and in so doing reduce the complexity of that system by identifying similar patterns that occur at different scales.  In many regards this process is the opposite of the models used for QCD particle production, which use a small number of basic rules, and from that generate a system with many degrees of freedom.    There are many different wavelet bases, but in this paper we shall use the Haar \cite{haar:hal-01333722} and Daubechies \cite{doi:10.1002/cpa.3160410705, Daubechies:1993:OBC:154993.155007} bases.

To enable the use of a discrete wavelet transform and the inverse transform, the event must first be converted to a two-dimensional pixel array. The pixel array covers the $y-\phi$ space of the event with a fixed number of pixels of equal size, each of which is assigned a value equal to the sum of the transverse momenta, \pT, of the particles emitted into the pixel. Following the method of \cite{Monk:2014uza}, a square pixel array of size $N\times N$ is chosen, and the two-dimensional discrete wavelet transformation is applied to that array. The dyadic hierarchy of the decompositon requires that the value of $N$ is an integer power of two. The transformation produces a set of wavelet coefficients, the number of which is equal to $N\times N$, the number of pixels in the array.  

The wavelet coefficients can be organised into an $N_{\mathcal{B}}\times N_{\mathcal{B}}$ array of frequency bands, labelled $\mathcal{B}_{\ly,\lphi}$.  The value of  $N_{\mathcal{B}}$ is given by equation \ref{eqn:NBands}.

\begin{equation}
  N_{\mathcal{B}} = \log_{2}\left(N\right) + 1
  \label{eqn:NBands}
\end{equation}

Frequency bands are labelled by two indices, $\ly$ and $\lphi$, that denote the angular scales in the $y$ and $\phi$ directions, respectively, to which the frequency band is sensitive.   We define the indices such that the frequency bands labelled by small values contain information on the large-scale details in the pixel array, while those frequency bands denoted by large values contain the small-scale details.  Thus frequency band $\mathcal{B}_{1,1}$ contains information on the low-frequency, wide-angle differences across the pixel array, while $\mathcal{B}_{7,7}$ contains the high-frequency, small-angle features.  Off-diagonal frequency bands, for example, $\mathcal{B}_{1,7}$, contain those structures that are small-angle in one direction, but wide-angle in another.  Such contributions may be of particular interest in studies of beam remnant connection effects in QCD.    

The wavelet coefficients in the frequency bands labelled by either $\ly = 0$ or $\lphi = 0$, or both, have a special significance and are called the smoothing coefficients.  The smoothing coefficients represent the lowest level of the decomposition and encode the widest angle features possible with the given wavelet basis.  In the case of the one dimensional Haar basis, the single smoothing coefficient is the simple average of all of the pixels in a column or row of $y$ or $\phi$.  The one dimensional Daubechies representation is less intuitive, but the four smoothing coefficients together encode the average activity in the array. 

Each frequency band is itself an array of wavelet coefficients of size $N_{\ly}\times N_{\lphi}$, with $N_{\ly}$ and $N_{\lphi}$ given by equation \ref{eqn:Nyphi}.

\begin{align}
N_{\ly} &= 
\begin{cases}
S_{w} & l_{y} = 0\\
2^{\left(l_{y} -1\right)}S_{w} & \text{otherwise}
\end{cases}\nonumber \\
N_{\lphi} &= 
\begin{cases}
S_{w} & l_{\phi} = 0\\
2^{\left(l_{\phi} - 1\right)}S_{w} & \text{otherwise}
\end{cases}\label{eqn:Nyphi}
\end{align}
where $S_{w}$ is the support of the wavelet basis used for the decomposition, which is determined by the complexity of the basis.  For the Haar basis, the support is $S_{w} = 1$, while for the Daubechies D4 basis $S_{w} = 4$.   The maximum valid values for $\ly$ and $\lphi$, $\ly^{\mathrm{max}}$ and $\lphi^{\mathrm{max}}$, are determined by the size of the pixel array, $N$, together with the support, $S_{w}$, and are constrained by the fact that the total number of coefficients is equal to the size of the pixel array, as in equation \ref{eqn:Nbands}.

\begin{align}
  \sum\limits_{\ly=0}^{\ly^{\mathrm{max}}}N_{\ly} &= N\nonumber \\
  \sum\limits_{\lphi=0}^{\lphi^{\mathrm{max}}}N_{\lphi} &= N\label{eqn:Nbands}
\end{align}

Thus a single frequency band is an array of wavelet coefficients that cover the entire $y - \phi$ plane and correspond to the same angular scales given by $l_{y}$ and $l_{\phi}$.  Frequency bands encode deviations from the mean activity at the angular scales to which the band corresponds.  Since the average deviation from the mean must be zero by definition, the average of the wavelet coefficient values within a single band must also be zero.  The high frequency contributions are encoded with more wavelet coefficients than the low frequency contributions, and the total number of wavelet coefficients in all of the frequency bands is equal to the total number of pixels in the original array.

\begin{figure}[t!]
  \captionsetup[subfigure]{singlelinecheck=off}
  \begin{subfigure}{0.5\textwidth}
    \flushleft
    \includegraphics[width=\textwidth]{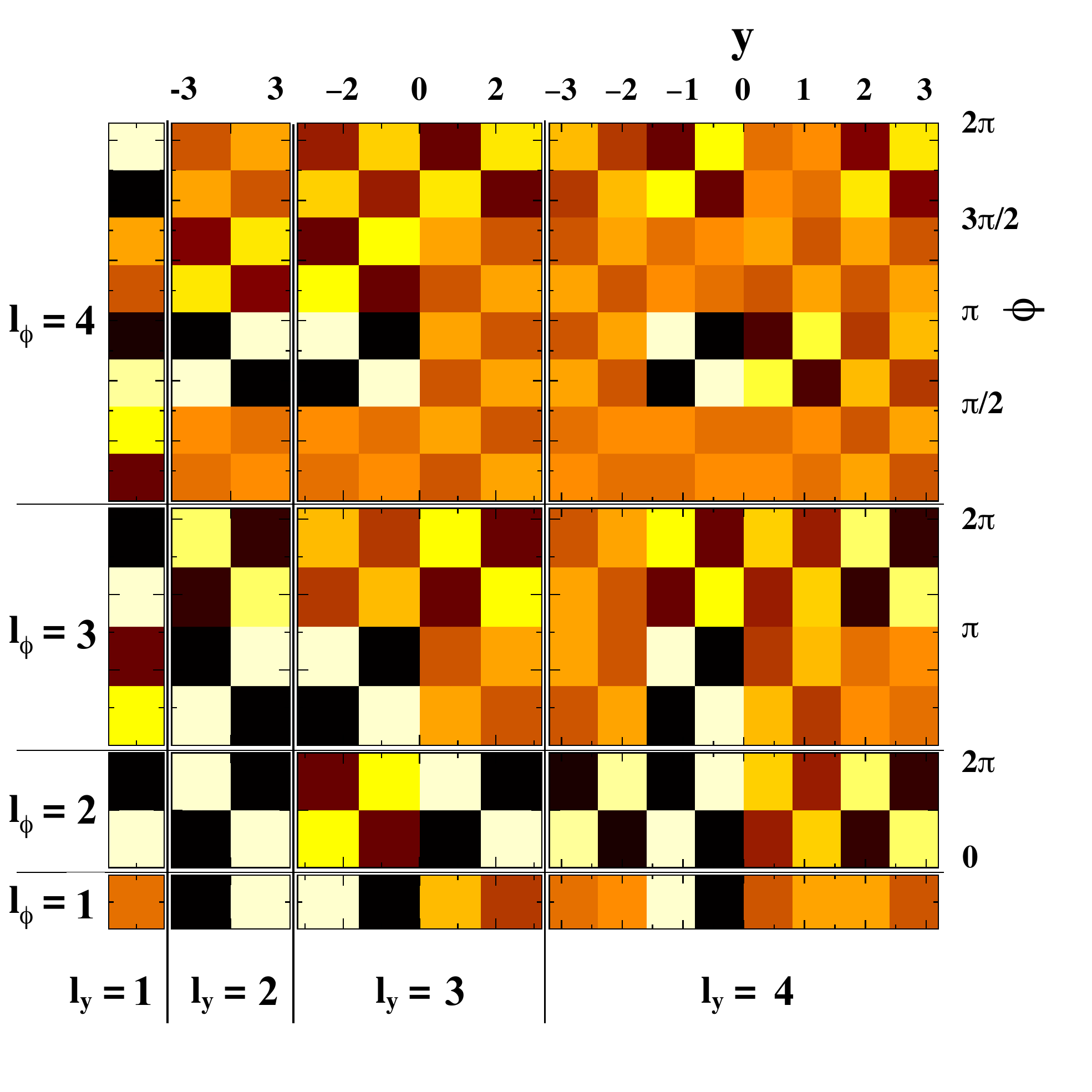}
    \vspace{-20pt}
    \caption{}
    \label{fig:levels_a}
  \end{subfigure}
  \begin{subfigure}{0.5\textwidth}
    \flushright
    \includegraphics[width=\textwidth]{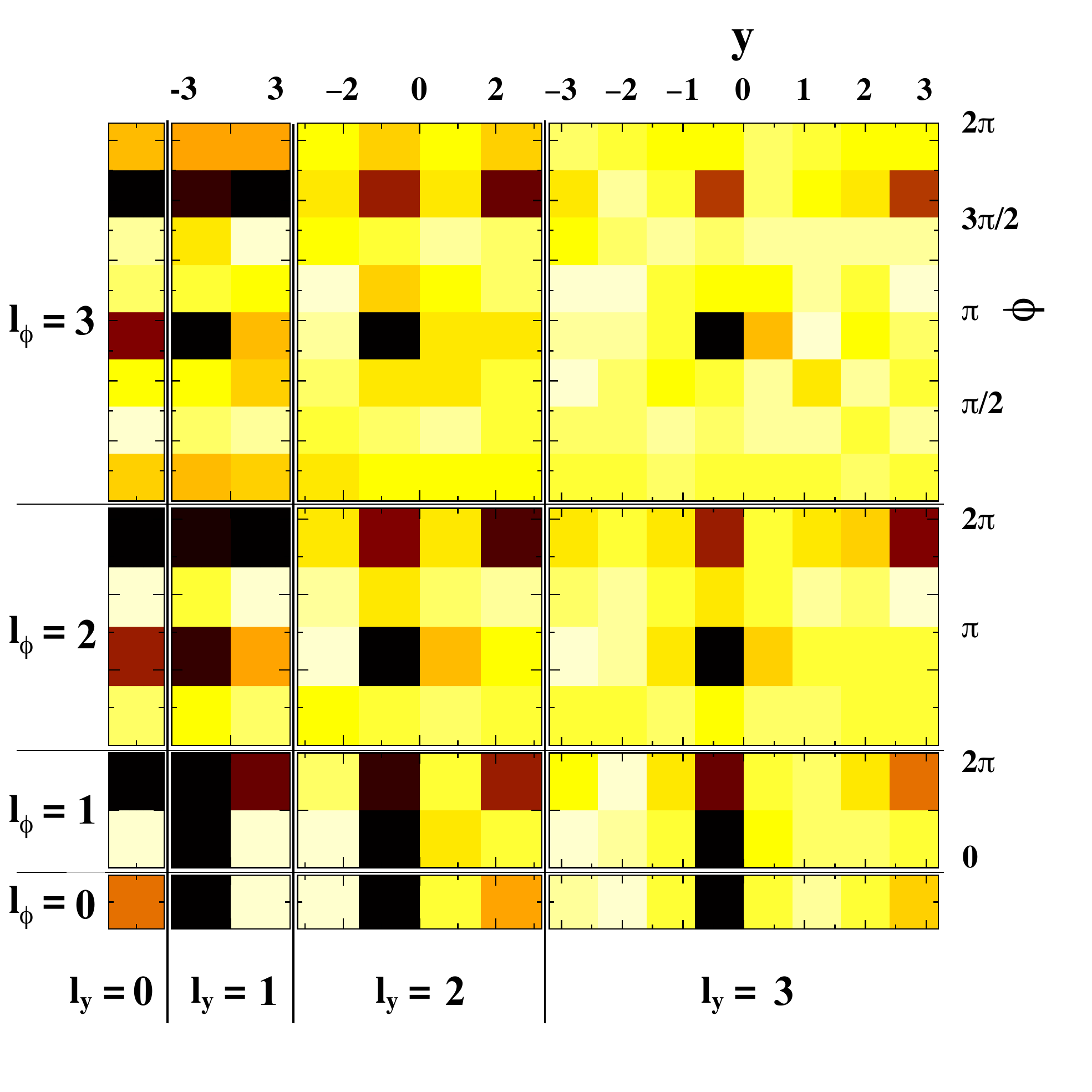}
    \vspace{-20pt}
    \caption{}
    \label{fig:levels_b}
  \end{subfigure}
  \caption{Multi-resolution view of a jet event using the Haar wavelet decomposition.  Each sub-panel of the left and right plots corresponds to wavelet frequency band, $\mathcal{B}_{l_{y},l_{\phi}}$, labelled on the bottom and left axes by $l_{y}$ and $l_{\phi}$, respectively.  Within each band, the $y, \phi$ coordinates are shown on the top and right axes, respectively.   Figure \ref{fig:levels_a} shows the individual wavelet coefficients in each frequency band, while Figure \ref{fig:levels_b} shows the activity in the event that is obtained when only a subset of the frequency bands is used to recover the event from the wavelet decomposition. Note that the bands containing smoothing coefficients are not shown in Figure \ref{fig:levels_a}. }
  \label{fig:levels}
\end{figure}

The effect of the hierarchical organisation of the wavelet coefficients as frequency bands is shown in Figure \ref{fig:levels}.  Figure \ref{fig:levels} uses a single Pythia 8 jet event, which has been decomposed using the Haar wavelet basis on a pixel array of size $N\times N = 16 \times 16$.   Figure \ref{fig:levels_a}, shows the size of the wavelet coefficients - expressed as colour intensity - that result from the wavelet decomposition.  Each sub-panel -  labelled by $\ly$ and $\lphi$ - corresponds to the single frequency band that matches those $\ly$ and $\lphi$ values.  Within each frequency band there is a set of coefficients that covers the entire $y - \phi$ space.  Sub-panels towards the bottom and left of the diagram show coarse wide-angle contributions, while the panel in the top right shows fine details.  

While  Figure \ref{fig:levels_a} shows the individual wavelet coefficients at different resolutions, Figure \ref{fig:levels_b} shows the event energy emission pattern that results from discarding different wavelet contributions from the event.   Each sub-panel in Figure \ref{fig:levels_b}, labelled by $\ly$ and $\lphi$, shows the energy distribution in the event when it is reconstructed with the  inverse wavelet transform using only those coefficients with wavelet indices equal to or larger than $\ly$ and $\lphi$. Thus, sub-panels towards the bottom and left of the diagram include only coarse wide-angle contributions, while the panel in the top right includes fine details. Off-diagonal sub-panels are able to resolve fine details in one co-ordinate, but only gross details in the other.   Panels are drawn on a smaller scale if they contain fewer coefficients, and therefore contain less information.

\section{Monte Carlo Event Samples}\label{sec:MCSamples}

Signal events are generated using Sherpa version 2.2 \cite{Gleisberg:2008ta, Schumann:2007mg, Krauss:2001iv, Gleisberg:2008fv, Hoeche:2009rj, Schonherr:2008av}. The signal process is $\nu\bar{\nu}$ production via a $Z$ boson, which provides an example of events in which there is missing transverse energy (\ETMiss) due to unobservable neutrinos. \ETMiss\ is known to be difficult to measure in the presence of pile-up. The default Sherpa tune with the NNPDF 3.0 \cite{Ball:2014uwa} PDF is used to generate 0.5 million proton-proton collision events with a centre-of-mass energy of 13~TeV.    

Pile-up is simulated using samples of soft QCD events generated with Pythia 8.205 \cite{Sjostrand:2006za, Sjostrand:2014zea}, which are overlaid on top of the signal events using the PileMC package \cite{pilemc}. The A2 Pythia tune \cite{ATLAS:2011krm} with the MSTW 2008 LO PDF \cite{Martin:2009iq} is used. Four separate samples of soft QCD events are generated: a diffractive sample plus three samples of non-diffractive events. The diffractive sample contains 100,000 events. The non-diffractive samples are labelled low \pT, medium-\pT\ and high \pT\ and are defined according to the \pT\ of the highest \pT\ parton in the event. The \pT\ requirement of the leading parton satisfies $p_{T} <$~15~GeV, 15~$\leq~p_{T} <$~50~GeV, or $p_{T} \geq$~50~GeV for the low, medium and high \pT\ samples, respectively. The low \pT\ sample contains 250,000 events, while the medium and high \pT\ samples each contain 100,000 events. 

The soft QCD events are combined to form a high pile-up sample by randomly selecting events from each of the four samples with a probability proportional to the sample cross-section. This allows high pile-up samples to be produced without generating prohibitively large numbers of soft QCD events. Three different pile-up scenarios are generated by overlaying $\mu$ events, where $\mu$ is a random number taken from Poisson distributions with means of $\left<\mu\right>=$~50, 100 and 300.

\section{Soft-QCD Collisions in the Wavelet Domain}
\label{sec:softqcd}

\begin{figure}[t!]
  \begin{subfigure}{0.5\textwidth}
    \centering
    \includegraphics[width=1.0\textwidth]{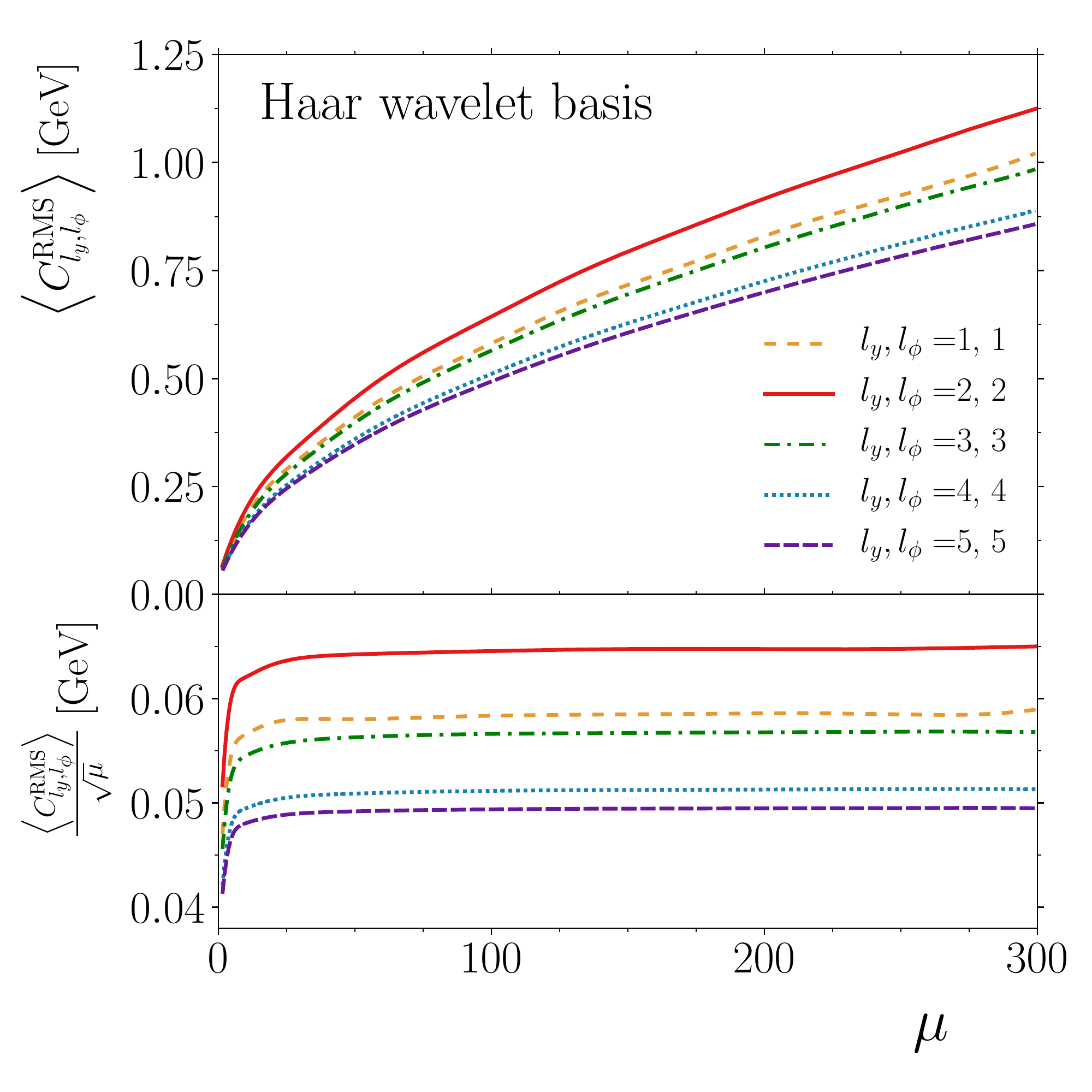}
  \end{subfigure}
  \begin{subfigure}{0.5\textwidth}
    \centering
    \includegraphics[width=1.0\textwidth]{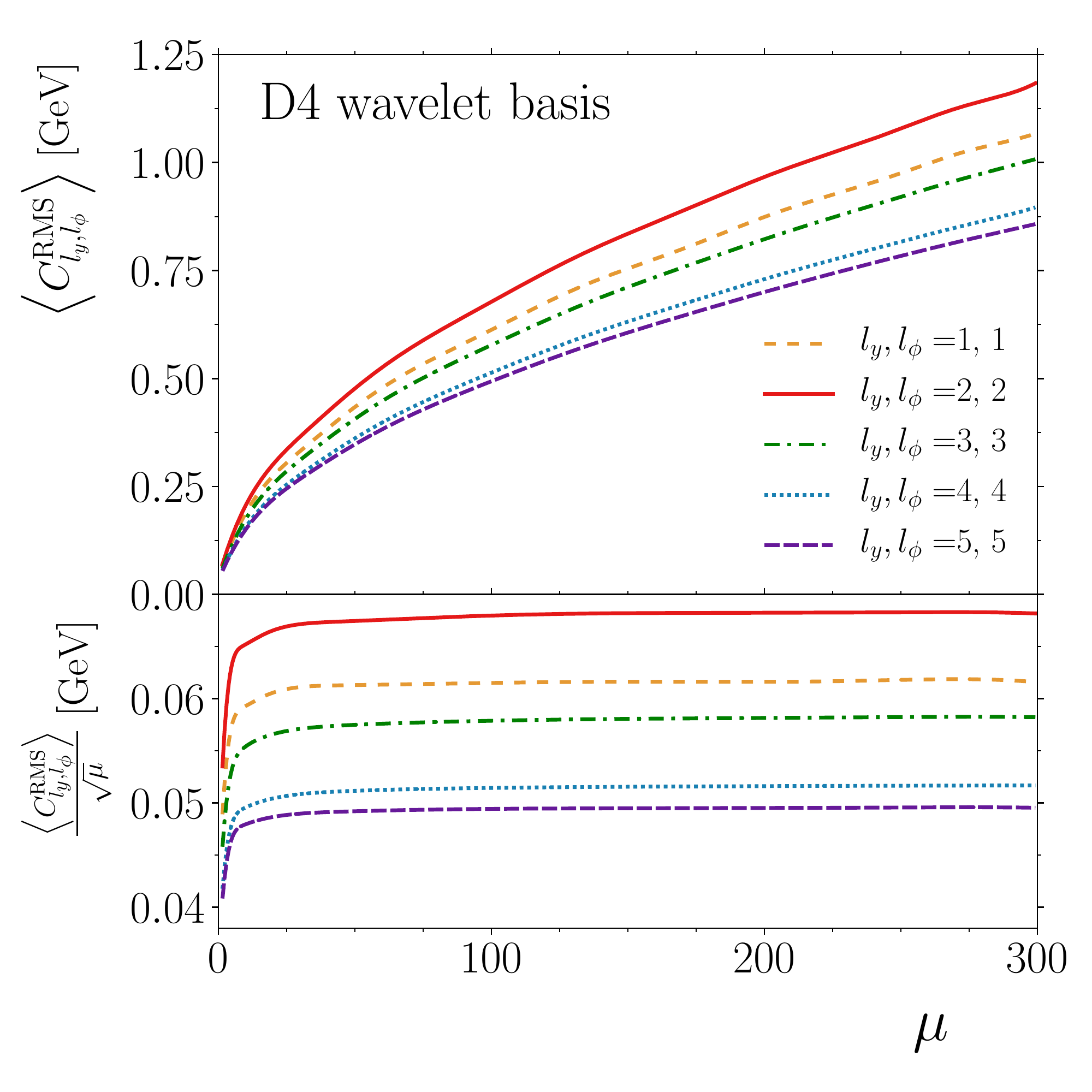}
  \end{subfigure}
  \caption{The dependence of the RMS of the diagonal frequency bands $\mathcal{B}_{\ly,\lphi}$ on $\mu$. The left- and right-hand plots use the Haar and D4 basis, respectively. The lower sub-panels of each plot show RMS/$\sqrt{\mu}$.}
  \label{fig:rms}
\end{figure}

In order to better understand the nature of pile-up and select an appropriate noise suppression scheme, we first study a sample of pile-up in the absence of any signal process.   Events are decomposed using either the Haar or the Daubechies D4 wavelet basis, and a pixel array size of $N\times N = 128\times 128$ covering the rapidity range $\left|y\right| < 3.2$ is used. A measure of the amount of activity that is occurring in an event when seen at different angular scales is the RMS of the wavelet coefficients in each frequency band, $\rms{\ly}{\lphi}$. For example, if $\rms{1}{1}$ is much larger than $\rms{7}{7}$ then the event has a lot of wide-angle activity, but not a lot of small-angle fluctuations. The dependence of some of the  $\rms{\ly}{\lphi}$ values on the number of overlaid soft inelastic collisions, $\mu$, is shown in Figure~\ref{fig:rms}. The $\mathcal{B}_{0,0}$ band always reflects the total \pT\ sum emitted into the event, which rises linearly with $\mu$, and for this reason is not shown. For all other frequency bands, $\rms{\ly}{\lphi}$ is proportional to $\sqrt{\mu}$ for values of $\mu\gtrsim 20$, although for brevity, only the diagonal frequency bands in which $\ly = \lphi$ are shown in Figure \ref{fig:rms}.  The Haar and Daubechies D4 wavelet bases are compared, and both show the same dependence on $\sqrt{\mu}$.


The dependence of $\rms{\ly}{\lphi}$ on $\sqrt{\mu}$ arises because the wavelet coefficients encode \emph{differences} in the \pT\ emitted into different regions, not the total amount of activity (which is encoded in the $\left\{0,0\right\}$ coefficient). These differences are Poissonially distributed fluctuations away from the average activity, and since the average activity must be proportional to $\mu$, the size of the fluctuations is proportional to $\sqrt{\mu}$. The deviation from this behaviour for $\mu$ values below around 20 occurs because the correlations within single proton-proton collisions have more significance at low $\mu$ than at high $\mu$, and the approximation that all emissions are uncorrelated ceases to be good.

\begin{figure}[t!]
  \centering
  \includegraphics[width=0.5\textwidth]{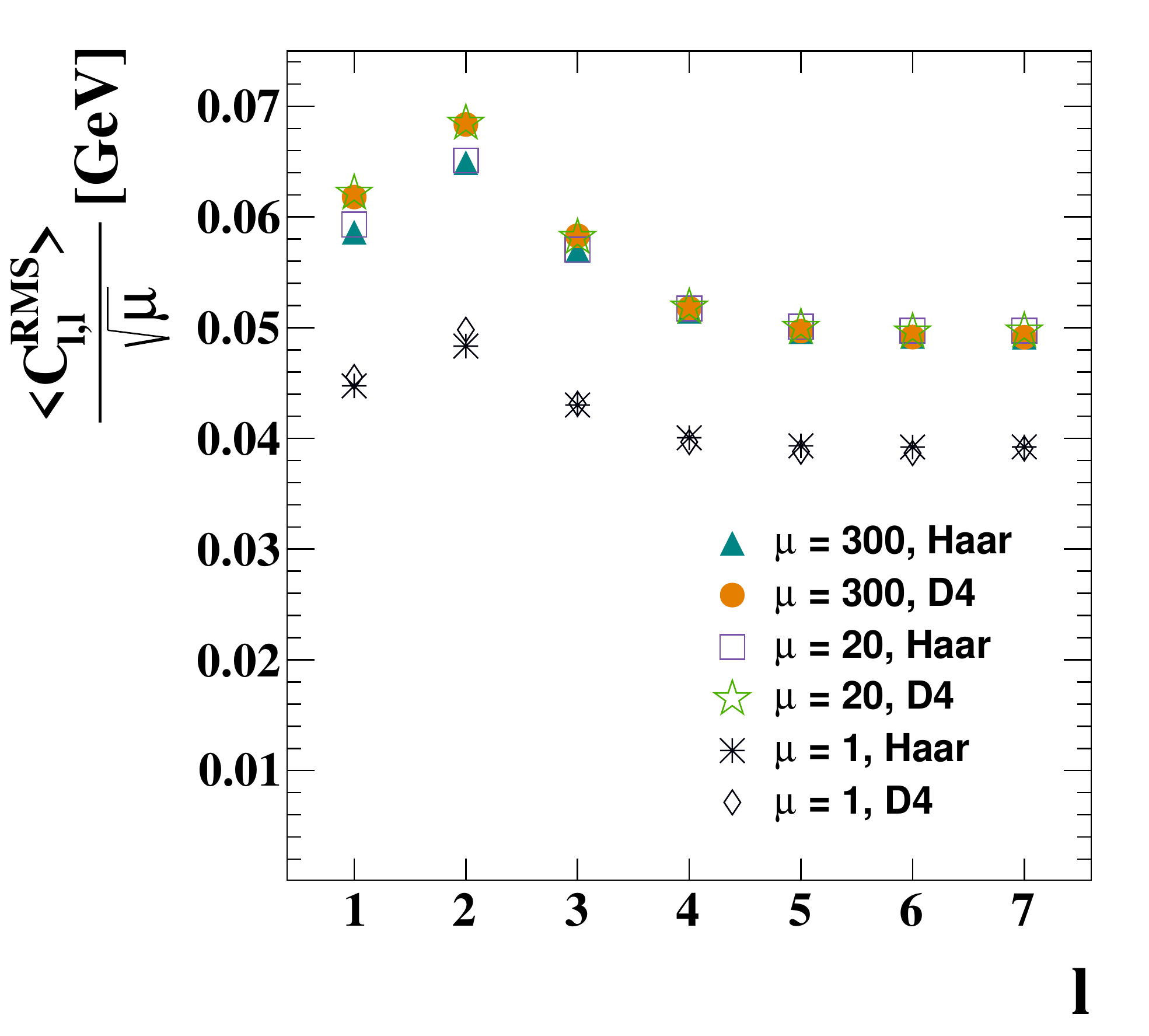}
  \caption[The RMS of diagonal frequency bands.] {The average value of the RMS of the diagonal frequency bands $\mathcal{B}_{l,l}$ (where $l = \ly = \lphi$), divided by $\sqrt{\mu}$. Results are shown when 1, 20 or 300 soft QCD events are overlaid.}
  \label{fig:levels_diagonal}
\end{figure}

The value of $\rms{l}{l}$ (where $l = \ly = \lphi$) averaged over 0.5 million events and divided by $\sqrt{\mu}$ is labelled \avrmsmu{l}{l} and is shown in Figure~\ref{fig:levels_diagonal} for all values of $l$ and three values of $\mu$. The Haar and the Daubechies D4 wavelet bases are also compared. There is very little difference between \avrmsmu{l}{l} at $\mu=20$ and $\mu=300$, regardless of the wavelet basis. Values of \avrmsmu{l}{l} have a small $l$ dependence, ranging between 0.05~GeV to 0.07~GeV. The lack of dependence on $l$ is consistent with the approximation that pile-up particles are uncorrelated. When there is only a single collision ($\mu=1$), the value of \avrmsmu{l}{l} is reduced by around 20\% compared to higher $\mu$ values.

The difference in behaviour at high and low $\mu$ is explored by looking at the asymmetry between the $\ly$ and $\lphi$ dependence of the wavelet coefficients. We define an asymmetry, $a_{\ly,\lphi}$, in equation~\ref{eqn:asymmetry}.

\begin{equation}
  a_{\ly,\lphi} = \frac{\rms{\ly}{\lphi} - \rms{\lphi}{\ly}}{\rms{\ly}{\lphi} + \rms{\lphi}{\ly}}
  \label{eqn:asymmetry}
\end{equation}

\begin{figure}[t!]
  \centering
  \includegraphics[width=0.5\textwidth]{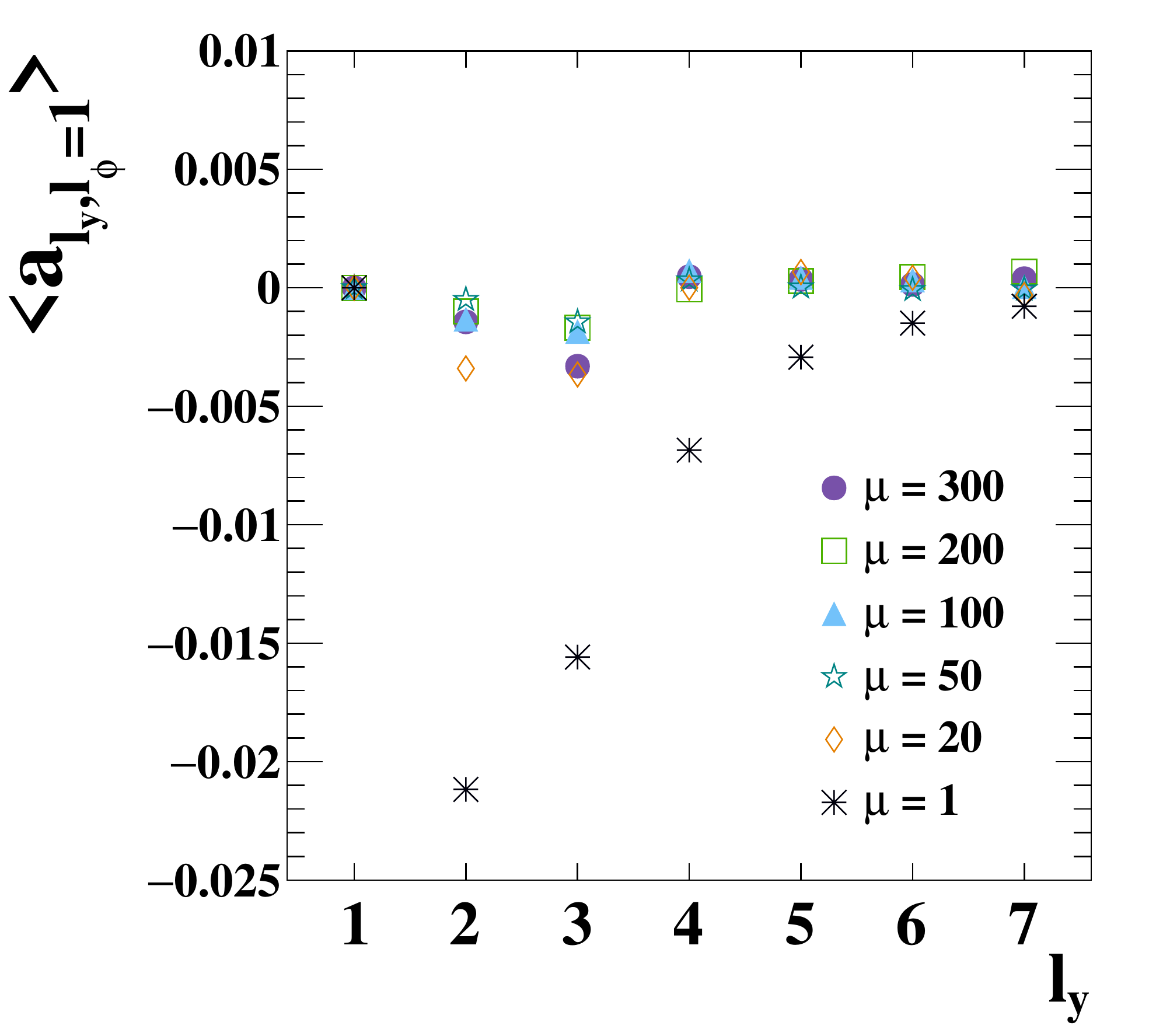}
  \caption[The asymmetry of frequency bands.] {The average asymmetry, $\langle a_{\ly,\lphi}\rangle$, between the strength of correlations in the $y$ and $\phi$ directions when decomposed using the Haar wavelet basis. $\lphi$ is fixed at one, and a scan of $\ly$ values is shown.}
  \label{fig:asymmetry}
\end{figure}

A large value for $a_{\ly,\phi}$ indicates that correlations between activity along the $y$ direction are different to correlations along the $\phi$ direction. Figure~\ref{fig:asymmetry} shows the mean value of $a_{\ly,\lphi}$ using the Haar wavelet basis for different values of $\ly$ when $\lphi=1$. Since small values of the $\ly$ or $\lphi$ level correspond to large scale structures, fixing $\lphi=1$ in Figure~\ref{fig:asymmetry} means that the asymmetry is sensitive to structures in the event that are large scale in the $\phi$ direction. When $\ly$ is also small, the asymmetry reveals differences between the long-range behaviour in the $y$ and $\phi$ directions, whereas when $\ly$ is large, the asymmetry reveals differences between large scale structures in the $\phi$ direction and small scale structures in $y$.

The average asymmetry for $\mu=1$ is negative at low $\ly$ and approaches zero at high $\ly$. This means there are differences in the large scale correlations in the $y$ and $\phi$ directions, but not differences in the small scale correlations, on average. The negative asymmetry indicates that, since $\lphi=1$, the frequency band corresponding to large scale structures in the $y$ direction is more active than the equivalent band that corresponds to large scale structures in the $\phi$ direction.   This negative asymmetry can be explained by beam connection effects and radiation from the initial state, which create long-range rapidity correlations. At higher $\mu$ values of 20, 50, 100, 200 and 300, the asymmetry is, to a good approximation, zero at all $\ly$ values. This lack of asymmetry demonstrates that the correlations within single collisions become negligible when pile-up is present and that the long range effects of QCD can no longer be seen as the collective particle emissions at high $\mu$ behave like white noise.

\section{Description of Thresholding Algorithms}\label{sec:description}

As outlined in Section \ref{sec:noise}, pile-up can be mitigated by applying a threshold to activity in the wavelet domain.  Having decomposed an event and obtained the set of wavelet coefficients, the noise threshold, $T$, of equation \ref{eqn:cprime} is applied to the coefficients, resulting in a set of modified coefficients. The inverse wavelet transformation is applied to the modified coefficients to produce a new pixel array that now includes the effect of applying the threshold in the wavelet domain. The ratio, $r_{i}$, is determined for the $i^{\mathrm{th}}$ pixel according to

\begin{equation}
  r_{i} = \frac{\tilde{p}_{i}}{p_{i}}
  \label{eqn:ri}
\end{equation}

where $p_{i}$ is the value of the $i^{\mathrm{th}}$ pixel in the array prior to the wavelet transformation, and $\tilde{p}_{i}$ is the value of the $i^{\mathrm{th}}$ pixel after the threshold is applied in the wavelet domain.

A value of $r$ close to unity indicates that the pixel has not been altered very much by the modifications made in the wavelet domain. On the other hand, a value of $r$ close to zero (or otherwise significantly different from unity) indicates that the pixel has reduced (or changed) in value as a result of the application of the threshold in the wavelet domain.  

Candidate pile-up particles are rejected according to the $r$ value of the pixel in which they lie. A threshold, $r_{cut}$, is chosen and any particle lying within a pixel for which $r _{i} < r_{cut}$ is removed from the event. The value of $r_{cut}$ is not critical, and a good first choice is around 0.7 for a wide range of pile-up conditions. The ratio $r$ has previously been used as a scaling factor in \cite{Monk:2014uza}; note that using $r$ as a scaling factor is an attempt to quantify the contribution of different - but interfering - processes to an emission, while using $r$ as a selection criterion labels physically separable pile-up and signal particles. The dependence of the method on $r_{cut}$ is studied in Section~\ref{sec:results}.   Two different thresholding algorithms are defined in Sections \ref{sec:flat} and \ref{sec:dynamic}.

\subsection{Simple Flat Wavelet Threshold}
\label{sec:flat}

The behaviour of soft QCD collisions in the wavelet domain, which is explored in Section~\ref{sec:softqcd}, is used to inform the simplest approach to the removal of pile-up by filtering the wavelet coefficients. Since the smoothing coefficients reflect the total amount of energy in the event, they are proportional to $\mu$. Therefore scaling the smoothing coefficients all by $1/\mu$ accounts for the average pile-up contribution per pixel. The vast majority of the wavelet coefficients are not smoothing coefficients, and instead describe the per-pixel fluctuations around the mean pixel energy.

The definition of the pile-up removal algorithm is as follows:

\begin{itemize}
\item Define a $N \times N$ pixel array, $A_{i}$, across $y$ and $\phi$, in which the value of each pixel is the sum of the \pTs of all of the particles that lie within it. $N$ should be as large as possible.
\item Perform a wavelet decomposition on the pixel array using the chosen basis. The resulting set of coefficients is $\left|C_{n}\right|$.
\item Modify $\left|C_{n}\right|$ by multiplying the smoothing coefficients by $1/\mu$. All other wavelet coefficients are filtered so that the coefficient value is set to zero if it satisfies $\left|C_{n}\right| \le \left(T \times\sqrt{\mu}\right)$.
\item Perform the inverse wavelet transform on the modified and filtered set of coefficients to obtain a filtered $N\times N$ pixel array, $F_{i}$. 
\item Divide the filtered pixel array, $F_{i}$, by the initial pixel array, $A_{i}$ to determine a pixel array of ratios, $r_{i}=F_{i}/A_{i}$.
\item Signal particles are selected by requiring that the pixel in which they lie satisfies $r_{i} \geq r_{cut}$. Any particle whose pixel fails the requirement is removed from the event.
\end{itemize}
 
Note that all of the observable inputs used in this algorithm are safe in the presence of soft or collinear splittings.

For this study we use $y_{\mathrm{max}}=3.2$ and N=128. The values of \avrmsmu{\ly}{\lphi} observed in Figure~\ref{fig:levels_diagonal} suggest the location of the pile-up noise floor is approximately $100$~MeV$\times\sqrt{\mu}$. In this particle-based study, we find that a threshold around $T=150$~MeV performs somewhat better than $T=100$~MeV; the higher value allows for the removal of more pile-up without unnecessarily degrading the signal. The value of the noise threshold may depend on the detection apparatus and should therefore be optimised in the context of the experimental conditions in which this method is used. Values of $r_{cut}$ can be chosen between 0 and 1. We have used $r_{cut}=0.7$; while the choice of $r_{cut}$ affects the amount of pile-up removed, the precise choice of $r_{cut}$ near 0.7 is not critical to the performance of the algorithm.

\subsection{Dynamic Wavelet Threshold}
\label{sec:dynamic}

All of the LHC detectors include an ability to measure high resolution tracks produced by the passage of charged particles. By resolving the production vertex of charged particles, it is possible to identify charged particles that do not originate from pile-up. The main challenge of pile-up removal by filtering wavelet coefficients is identifying regions in the wavelet domain that contain signal contributions, and consequently not removing activity in those regions. The existence of known charged signal particles can aid in the identification of signal regions and thus improve the separation between signal and noise.

An improved algorithm using tracking information performs an additional wavelet decomposition on charged particles that are known to originate from the signal collision. The wavelet coefficients, $C_{n}^{\mathrm{trk}}$, that are obtained from the signal charged particle tracks can be used to provide a dynamic threshold for filtering. The threshold should be low for coefficients that are known to overlap with the charged signal, and high for regions that do not overlap the charged part of the signal.

A suitable ad-hoc threshold, $t\left(n\right)$, for the $n^{th}$ coefficient is given by equation~\ref{eqn:threshold}

\begin{equation}
  \frac{t\left(n\right)}{\sqrt{\mu}} = \left(1 - \erf(\frac{\left|C_{n}^{\mathrm{trk}}\right| -0.06 \mathrm{GeV} }{0.025 \mathrm{GeV}})\right)\times0.4\mathrm{~GeV}
  \label{eqn:threshold}
\end{equation}


\begin{figure}[t!]
  \centering
  \includegraphics[width=0.5\textwidth]{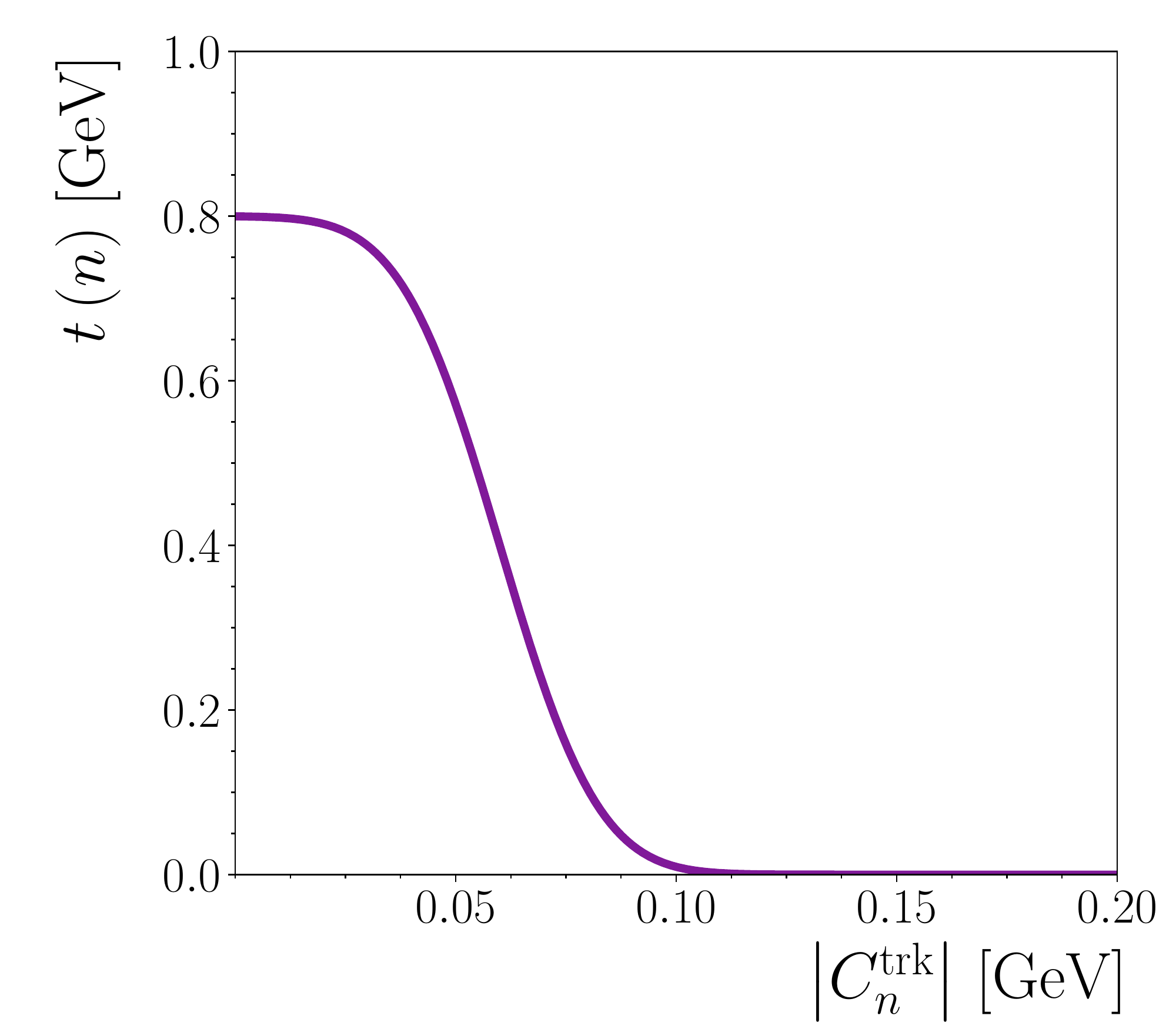}
  \caption[Turnoff filter threshold] {The filtering threshold used for the $n^{th}$ coefficient, $t\left(n\right)$, as a function of the corresponding charged particle wavelet coefficient, $C_{n}^{\mathrm{trk}}$.}
  \label{fig:turnoff}
\end{figure}

which is related to a Gaussian function with a width of $0.025$~GeV and a mean of $0.06$~GeV, and where $C_{n}^{\mathrm{trk}}$ is the $n^{\mathrm{th}}$ wavelet coefficient for charged particle tracks originating from the signal vertex. Note that the dependence on $\sqrt{\mu}$ is the same as in the "flat" case of Section~\ref{sec:flat} and that equation~\ref{eqn:threshold} should therefore be multiplied by the value of $\sqrt{\mu}$ for a given event. The "turnoff" curve described by equation~\ref{eqn:threshold} is shown in Figure~\ref{fig:turnoff}. In comparison to the flat-threshold of Section~\ref{sec:flat}, Figure~\ref{fig:turnoff} shows a higher threshold that approaches 0.8~GeV in regions of wavelet space lacking any charged signal activity (ie. in the limit $\left|C_{n}^{\mathrm{trk}}\right| \rightarrow 0$), and a threshold close to zero in regions of wavelet space that contain charged signal activity.

Since the charged particle tracking acceptance of the LHC detectors typically only covers the pseudo-rapidity, $\eta$, range $\left|\eta\right|<2.5$, the threshold used in the region between $\left|\eta\right| > 2.5$ and $\left|y\right| < 3.2$ must be interpolated between the flat-threshold described in Section~\ref{sec:flat} and the track-based threshold of equation~\ref{eqn:threshold}. The different wavelet coefficients cover different rapidity spans; some coefficients correspond to regions contained within $\left|\eta\right| < 2.5$, while others are partialy or entirely outside of $\left|\eta\right| < 2.5$. The region that each coefficient covers can be inferred from its $\ly$ index, which corresponds to its scale in the $y$ co-ordinate, together with the positional index. Together, these imply a minimum and maximum for the coefficient's rapidity coverage, $y_{\mathrm{min}}$ and $y_{\mathrm{max}}$, respectively. Taking the noise threshold from equation~\ref{eqn:threshold} for $\left|\eta\right|<2.5$, and the flat-threshold of $0.15\times\sqrt{\mu}$~GeV for $2.5 < \left|\eta\right| < 3.2$ and interpolating for each coefficient gives equation~\ref{eqn:overlap}, which describes the threshold $\tau\left(n\right)$ that is interpolated to account for the region beyond $\left|\eta\right| > 2.5$.

\begin{align}
  \Delta y_{\mathrm{max}} &= \begin{cases} y_{\mathrm{max}} - 2.5, & \text{if  } y_{\mathrm{max}} > 2.5\\
                                                                0, & \mathrm{otherwise}
                                                                \end{cases} \nonumber \\
\Delta y_{\mathrm{min}} &= \begin{cases} -2.5 -y_{\mathrm{max}}, & \text{if  } y_{\mathrm{max}} < -2.5\\
                                                                0, & \mathrm{otherwise}
                                                                \end{cases} \nonumber \\
\beta &= \begin{cases} 1 - \frac{\left(\Delta y_{\mathrm{max}} + \Delta y_{\mathrm{min}}\right)}{y_{\mathrm{max}} - y_{\mathrm{min}}}, & \text{if } \frac{\left(\Delta y_{\mathrm{max}} + \Delta y_{\mathrm{min}}\right)}{y_{\mathrm{max}} - y_{\mathrm{min}}} < 1\\
                                                                0, & \mathrm{otherwise}
                                                                \end{cases} \nonumber \\
\frac{\tau\left(n\right)}{\sqrt{\mu}} &= \beta\times \frac{t\left(n\right)}{\sqrt{\mu}} + \left(1-\beta\right)\times 0.15~\mathrm{GeV}\label{eqn:overlap}
\end{align}

In addition to allowing a more aggressive threshold on wavelet coefficients, the charged particle track information can also be used to provide a better scaling factor for the smoothing coefficients. The ratio, $S$, of the sums of charged particle \pTs shown in equation~\ref{eqn:scaling}

\begin{equation}
  S = \frac{\sum\limits_{i \in \mathrm{\scriptstyle{signal}}}p_{T i}^{trk} }{ \sum\limits_{i \in \mathrm{all}}p_{T i}^{trk}}\label{eqn:scaling}
\end{equation}

provides an estimate of the factor by which the smoothing coefficients should be scaled in order to account for the increase in total activity caused by pile-up.  

Finally, a weak $\mu$ dependence can be introduced to $r_{\mathrm{cut}}$. At very high pile-up levels, a more aggressive selection criteria using a higher value of $r_{\mathrm{cut}}$ is preferred, which removes more of the pile-up particles at the expense of removing somewhat more signal. Conversely, at low pile-up, a less stringent selection using a lower value of $r_{\mathrm{cut}}$ can help to preserve more of the signal. As with all pile-up mitigation schemes, it is harder to separate low \pT particles originating from pile-up and signal than it is to separate high \pT particles, which necessarily results in the retention and removal of some soft pile-up and signal particles, respectively. The optimal choice of $r_{\mathrm{cut}}$ will ideally provide a balance between retaining soft signal particles and rejecting soft pile-up particles. Since pile-up increases with $\mu$, this suggests that $r_{\mathrm{cut}}$ should also increase with $\mu$. An in-depth study of $r_{\mathrm{cut}}$ in an experimental context will reveal the appropriate $\mu$ dependence, but we choose the value of $r_{\mathrm{cut}}\left(\mu\right)$ given in equation~\ref{eqn:rcut}

\begin{equation}
  r_{\mathrm{cut}}\left(\mu\right) = 1 - \frac{1}{1+0.3\sqrt{\mu}}\label{eqn:rcut}
\end{equation}

Together with the new threshold, the charged particle track based algorithm is thus

\begin{itemize}
\item{Use all stable visible final state charged particles from the signal collision that satisfy $\left|\eta\right| < 2.5$. }
\item{Define a $N\times N$ pixel array, $M_{i}$ within $\left|y\right|<3.2$ in which the value of each pixel is the sum of the \pTs of all of the visible signal charged particles that lie within it.  As with the flat-threshold of Section \ref{sec:flat}, $N=$128 is used.}
\item{Perform a wavelet decomposition on the charged particle pixel array $M_{i}$ using the chosen basis.  The resulting set of coefficients is $C_{n}^{\mathrm{trk}}$.}
\item{Use all stable visible final state particles from the combined signal and pile-up events that satisfy $\left|y\right|<3.2$.}
\item{Define a $N\times N$ pixel array, $A_{i}$ within $\left|y\right|<3.2$ in which the value of each pixel is the sum of all the visible particles that lie within it.}
\item{Perform a wavelet decomposition on $A_{i}$.  The resulting set of wavelet coefficients is $C_{n}$.}
\item{Modify the smoothing coefficients of $C_{n}$ by multiplying them by $S$, the ratio of the sums of track \pTs from equation \ref{eqn:scaling}.}
\item{For each of the wavelet coefficients $C_{n}$ that are not smoothing coefficients, identify the corresponding coefficient from $C_{n}^{\mathrm{trk}}$ and determine a dynamic threshold, $\tau\left(n\right)$ by using equations \ref{eqn:threshold}  and \ref{eqn:overlap} with $C_{n}^{\mathrm{trk}}$.}
\item{Filter $C_{n}$ by setting its value to zero if $\left|C_{n}\right| \leq \tau\left(n\right)$}
\item{Perform the inverse wavelet transform on the modified and filtered set of coefficients, $C_{n}$, to obtain a filtered $N\times N$ pixel array, $A_{f}$}
\item{Divide the filtered pixel array, $A_{f}$, by the initial pixel array, $A_{i}$ to determine a pixel array of ratios, $r_{i}=A_{f}/A_{i}$. }
\item{Find which pixel $r_{i}$ each visible particle lies within.  If the value of the pixel satisfies $r_{i} < r_{cut}\left(\mu\right)$ then remove every particle that lies within that pixel.  The value of $r_{cut}$ is dependent on $\mu$ and is taken from equation \ref{eqn:rcut}.}
\end{itemize}

\section{Results}
\label{sec:results}

Both of the algorithms described in Section~\ref{sec:description} make a decision as to whether each particle in an event should be labelled as a pile-up particle or a signal particle. Since, in a Monte Carlo sample, the true origin of each particle is known, signal efficiencies and pile-up rejection rates can be determined for various samples and \avnp\ values.  We use the Rivet framework \cite{Buckley:2010ar} to perform analysis on the Monte Carlo samples described in Section \ref{sec:MCSamples}. The wavelet decomposition is performed using the NewWave package \cite{newwave, Monk:2014uza}.

The algorithms described in Section~\ref{sec:description} assign each particle in an event an $r$-value, and the particle is kept or discarded depending on whether its $r$-value is above or below the chosen $r_{\mathrm{cut}}$ value. The distribution of $r$ values from particles produced in \Znunu\ with \avnp~=~100 when using the flat-threshold algorithm (Section~\ref{sec:flat}) and the Haar wavelet basis is shown in Figure~\ref{fig:rvalues}. The distribution of $r$ values is normalised to unit area, and the contributions from particles originating from the signal collisions and pile-up collisions are shown separately. The $r$ value distribution is shown inclusively for all particles in Figure~\ref{fig:rvalues}(\subref{fig:rvalues_a}), and for particles that satisfy \pT~$<$~1~GeV, 1~$<$~\pT~$<$~4~GeV, and \pT~$>$~4~GeV in Figure~{\ref{fig:rvalues}(\subref{fig:rvalues_b}), (\subref{fig:rvalues_c}) and (\subref{fig:rvalues_d}), respectively.

\begin{figure}[t!]
  \captionsetup[subfigure]{singlelinecheck=off}
  \begin{subfigure}{0.5\textwidth}
    \centering
    \includegraphics[width=1.0\textwidth]{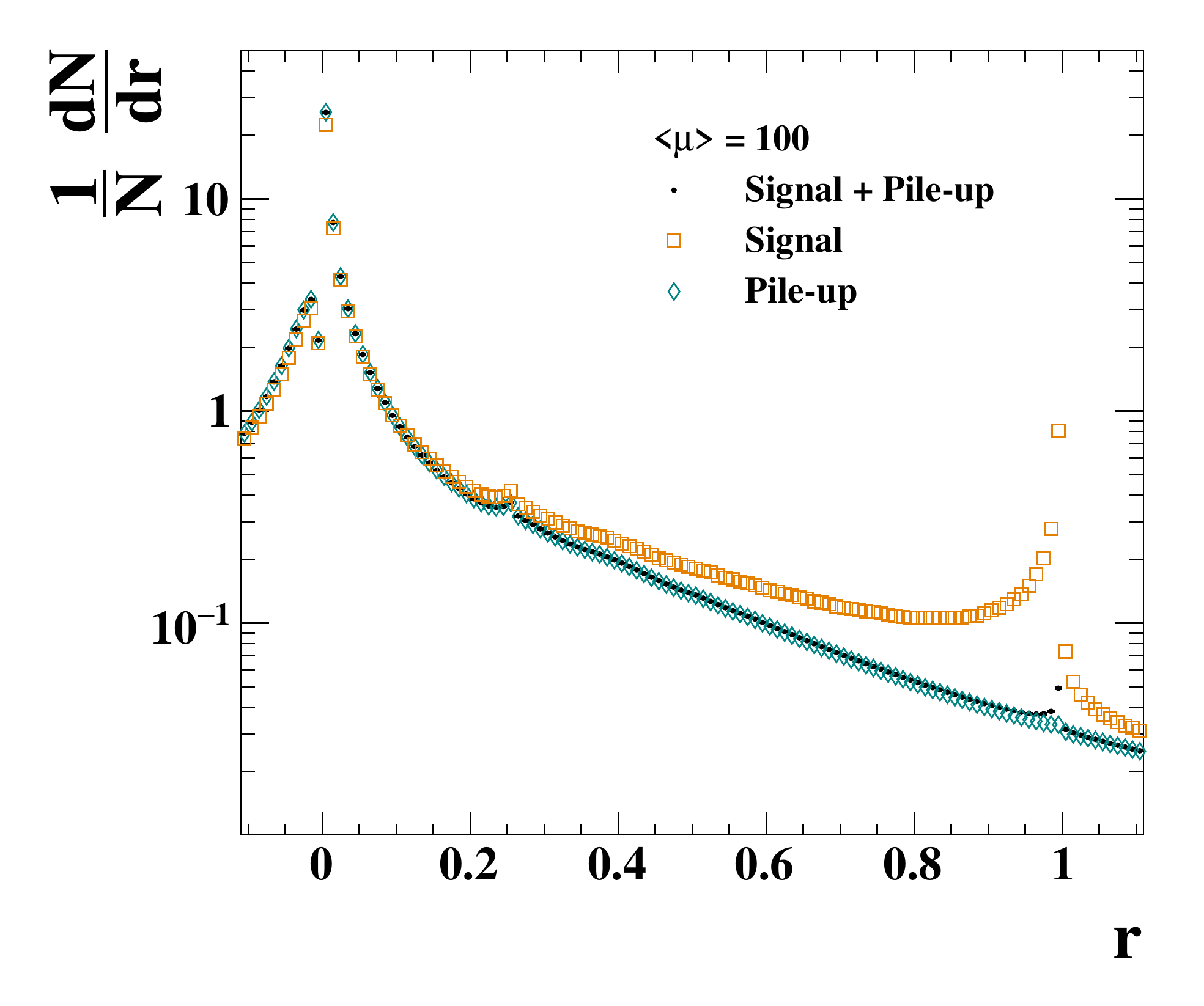}
    \vspace{-30pt}
    \caption{}
    \label{fig:rvalues_a}
  \end{subfigure}
  \begin{subfigure}{0.5\textwidth}
    \centering
    \includegraphics[width=1.0\textwidth]{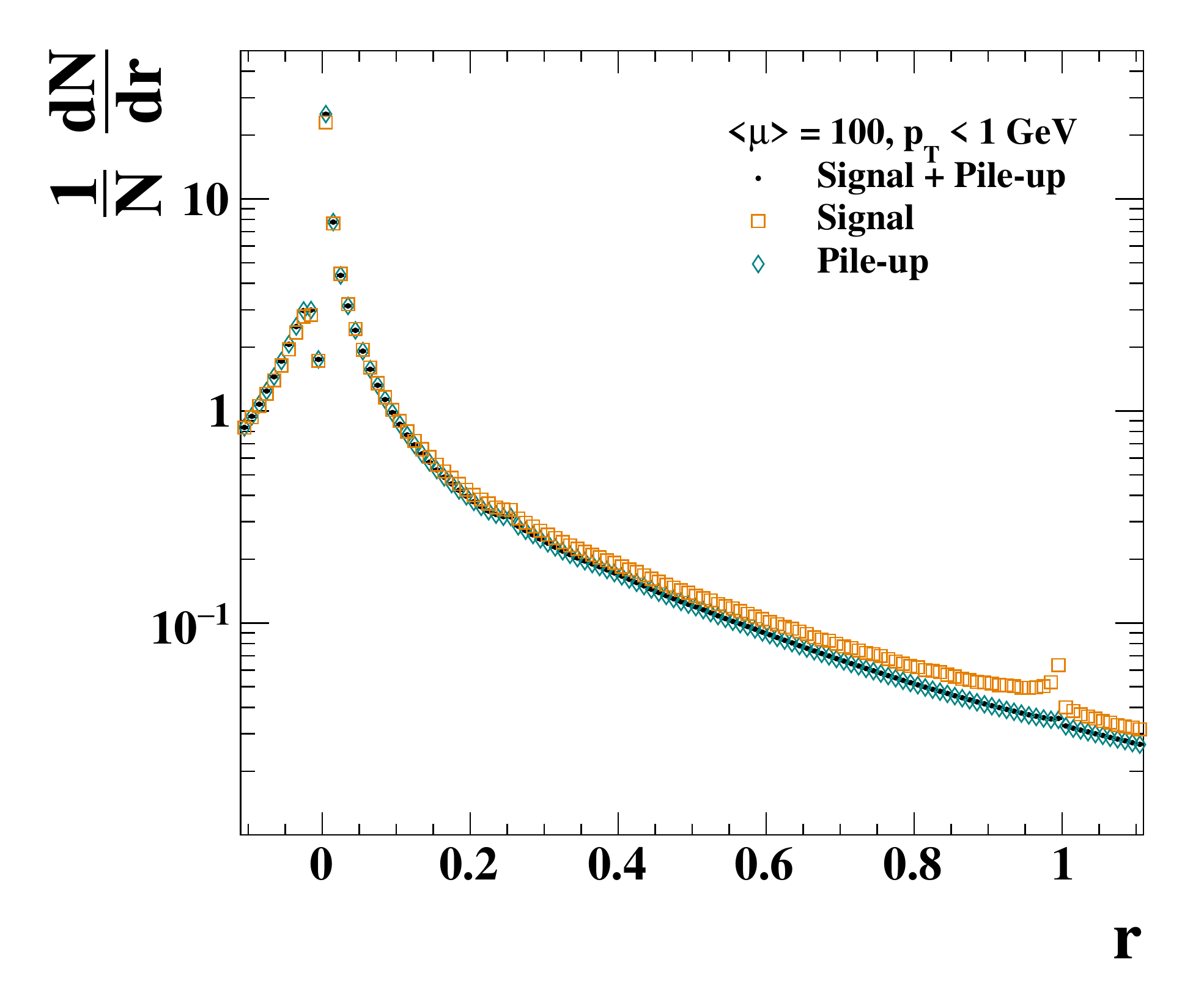}
    \vspace{-30pt}
    \caption{}
    \label{fig:rvalues_b}
  \end{subfigure}
  \begin{subfigure}{0.5\textwidth}
    \centering
    \includegraphics[width=1.0\textwidth]{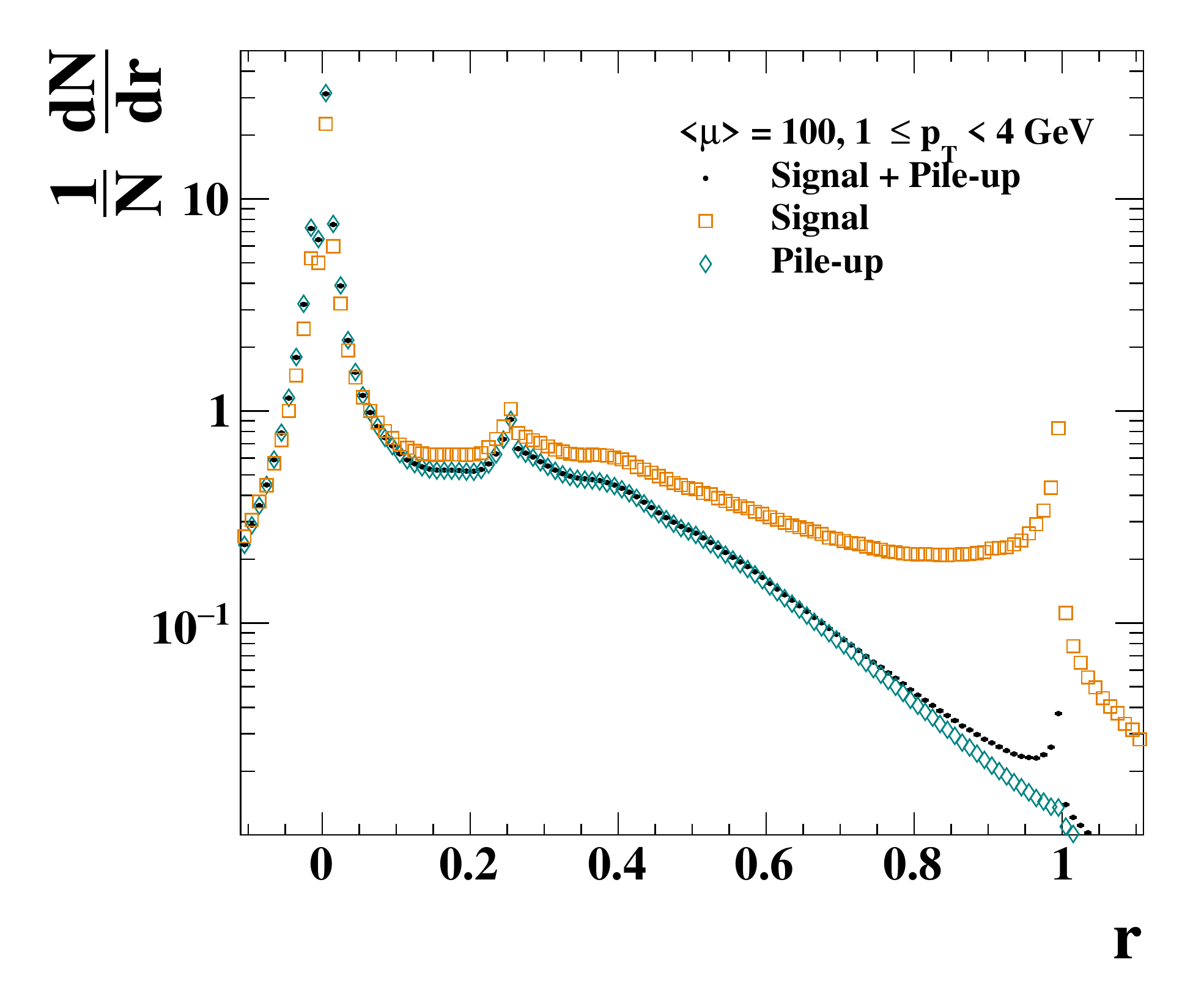}
    \vspace{-30pt}
    \caption{}
    \label{fig:rvalues_c}
  \end{subfigure}
  \begin{subfigure}{0.5\textwidth}
    \centering
    \includegraphics[width=1.0\textwidth]{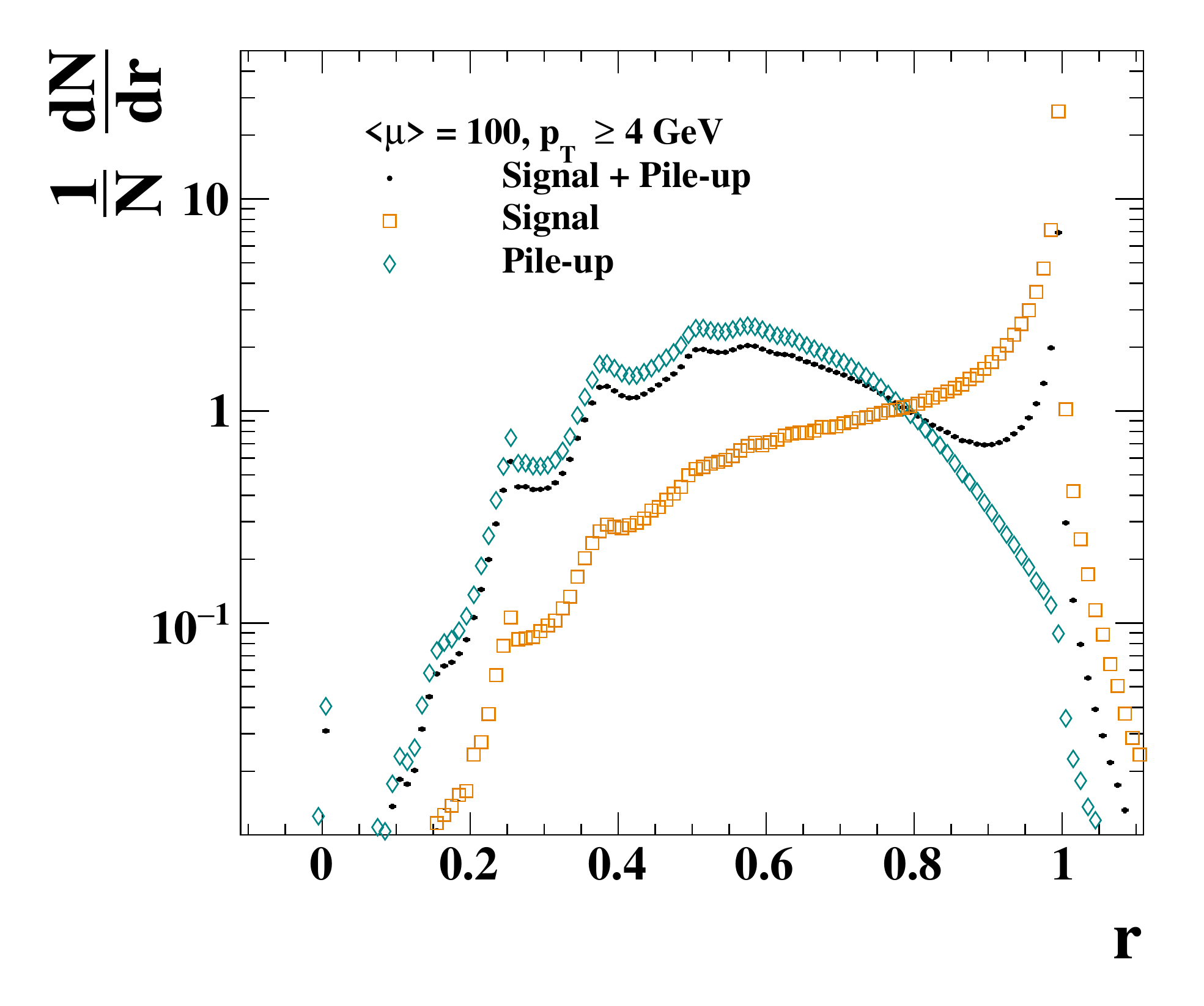}
    \vspace{-30pt}
    \caption{}
    \label{fig:rvalues_d}
  \end{subfigure}
  \caption[Distribution of r values] {Distribution of particle r values in \Znunu\ events with \avnp~=~100 when using the flat-threshold algorithm (Section~\ref{sec:flat}) and the Haar wavelet basis. The distribution of $r$-values, normalised to unit area, is shown for all particles (\subref{fig:rvalues_a}), and for those particles that satisfy \pT~$<$~1~GeV, 1~$<$~\pT~$<$~4~GeV, and \pT~$>$~4~GeV in (\subref{fig:rvalues_b}), (\subref{fig:rvalues_c}) and (\subref{fig:rvalues_d}), respectively.}
  \label{fig:rvalues}
\end{figure}

Figure~\ref{fig:rvalues}(\subref{fig:rvalues_a}) shows that particles originating from the signal collision tend to have $r$ values closer to one than particles originating from pile-up collisions, which have $r$ values closer to zero. It is much easier to distinguish signal and pile-up particles when the particle \pT\ is high, which is demonstrated by the well separated signal and pile-up peaks in Figure~\ref{fig:rvalues}(\subref{fig:rvalues_d}). At low \pT, the signal and pile-up distributions overlap to a greater extent, and it is harder to distinguish the two, as shown by Figure~\ref{fig:rvalues}(\subref{fig:rvalues_b}).

Low \pT\ particles are less important for the correct measurement of event observables than higher \pT\ particles. In order to give lower weight to such low \pT\ particles, an $r$ distribution in which each particle enters the histogram with a weight equal to its \pT\ is shown in Figure~\ref{fig:rByPT}. Figure~\ref{fig:rByPT} illustrates how much energy is emitted at a given $r$ value per event, and exhibits a good separation between signal contributions and pile-up contributions. 

\begin{figure}[t!]
  \centering
  \includegraphics[width=0.5\textwidth]{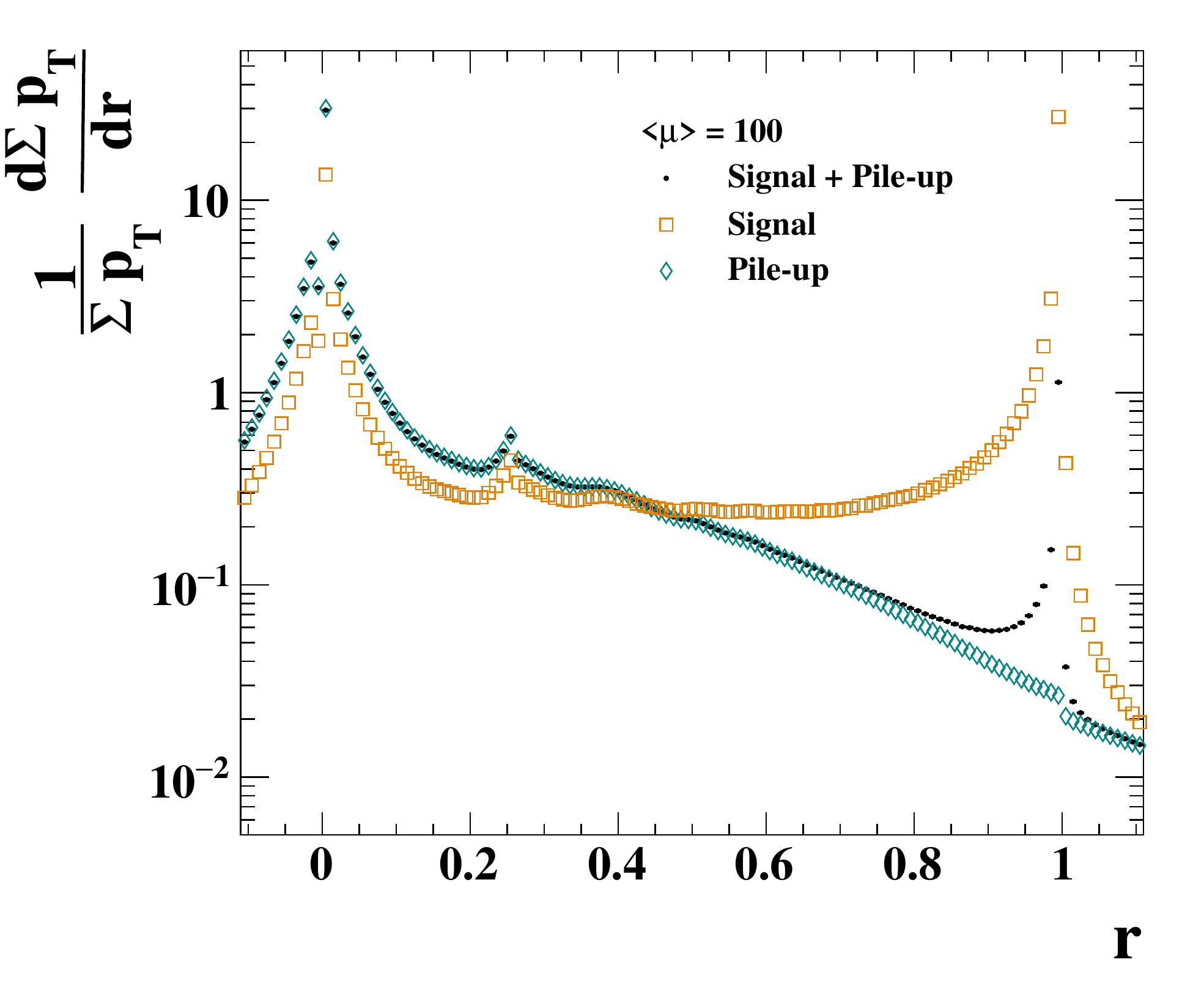}
  \caption[Distribution of r values weighted by \pT] {Distribution of particle r values in \Znunu\ events with \avnp~=~100 when using the flat-threshold algorithm (Section~\ref{sec:flat}) and the Haar wavelet basis. Each particle enters the distribution with a weight equal to its \pT.}
  \label{fig:rByPT}
\end{figure}

An improved separation between signal and pile-up particles is seen when using the dynamic-threshold algorithm (Section~\ref{sec:dynamic}) and the Haar wavelet basis, as shown in Figure~\ref{fig:rtracks}. Figure~\ref{fig:rvalues}(\subref{fig:rvalues_d}) already showed that the separation between signal and pile-up is good for high \pT\ particles, and the addition of the dynamic threshold using charged tracks extends the kinematic range so that good performance is seen at lower particle \pT.  This improvement in low-\pT\ performance due to charged particle tracking is revealed by comparing the $r$ distribution for \pT~$<$~1~GeV in Figure~\ref{fig:rtracks}(\subref{fig:rtracks_b}) with Figure~\ref{fig:rvalues}(\subref{fig:rvalues_b}).  Furthermore, when using the dynamic threshold algorithm, the $r$ distribution weighted by \pT\ also shows an improved separation between signal and pile-up, which can be seen by comparing Figure~\ref{fig:rByPT} with \ref{fig:rtracks}(\subref{fig:rtracks_a}).

\begin{figure}[t!]
  \captionsetup[subfigure]{singlelinecheck=off}
  \begin{subfigure}{0.5\textwidth}
    \centering
    \includegraphics[width=1.0\textwidth]{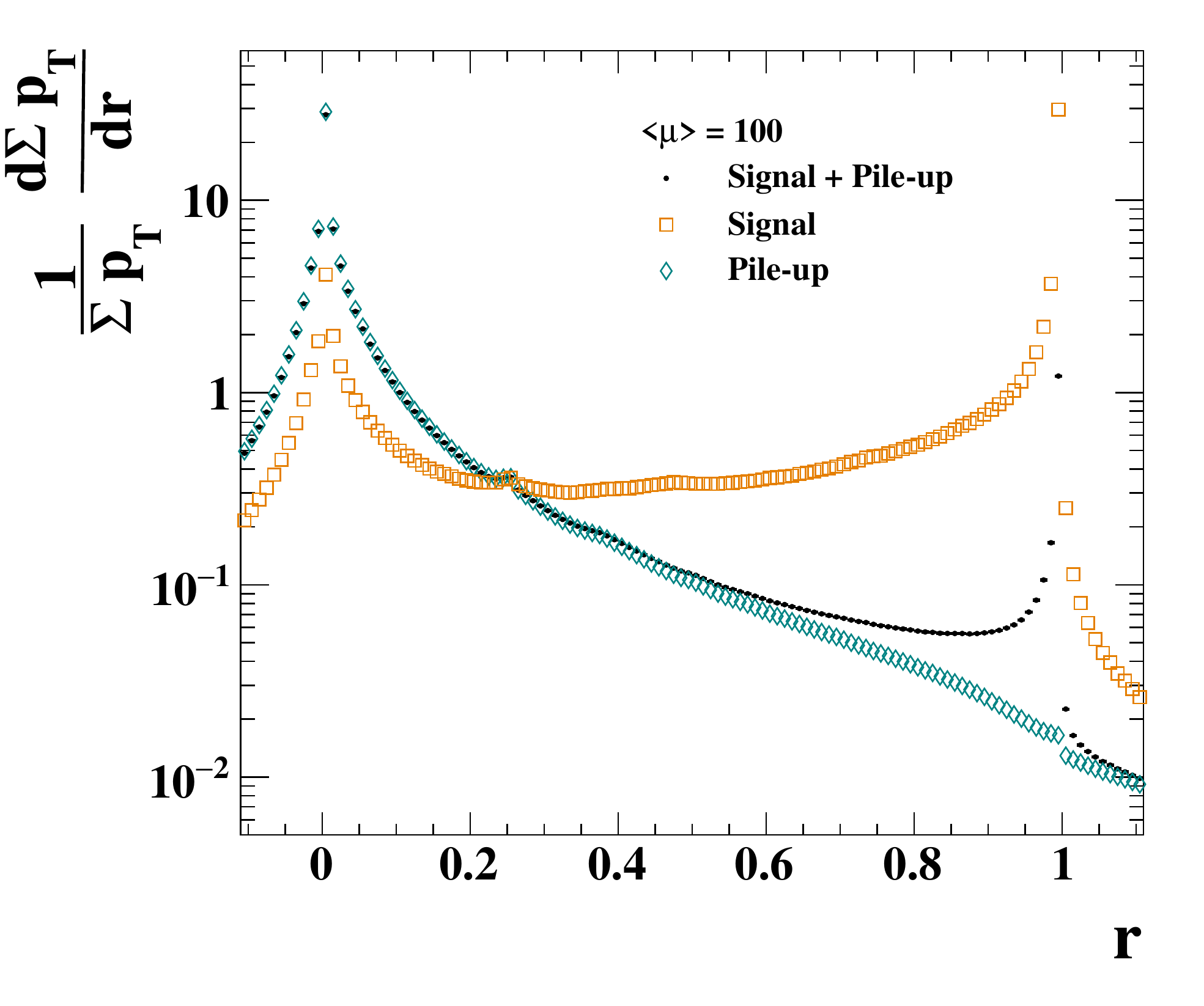}
    \vspace{-30pt}
    \caption{}
    \label{fig:rtracks_a}
  \end{subfigure}
  \begin{subfigure}{0.5\textwidth}
    \centering
    \includegraphics[width=1.0\textwidth]{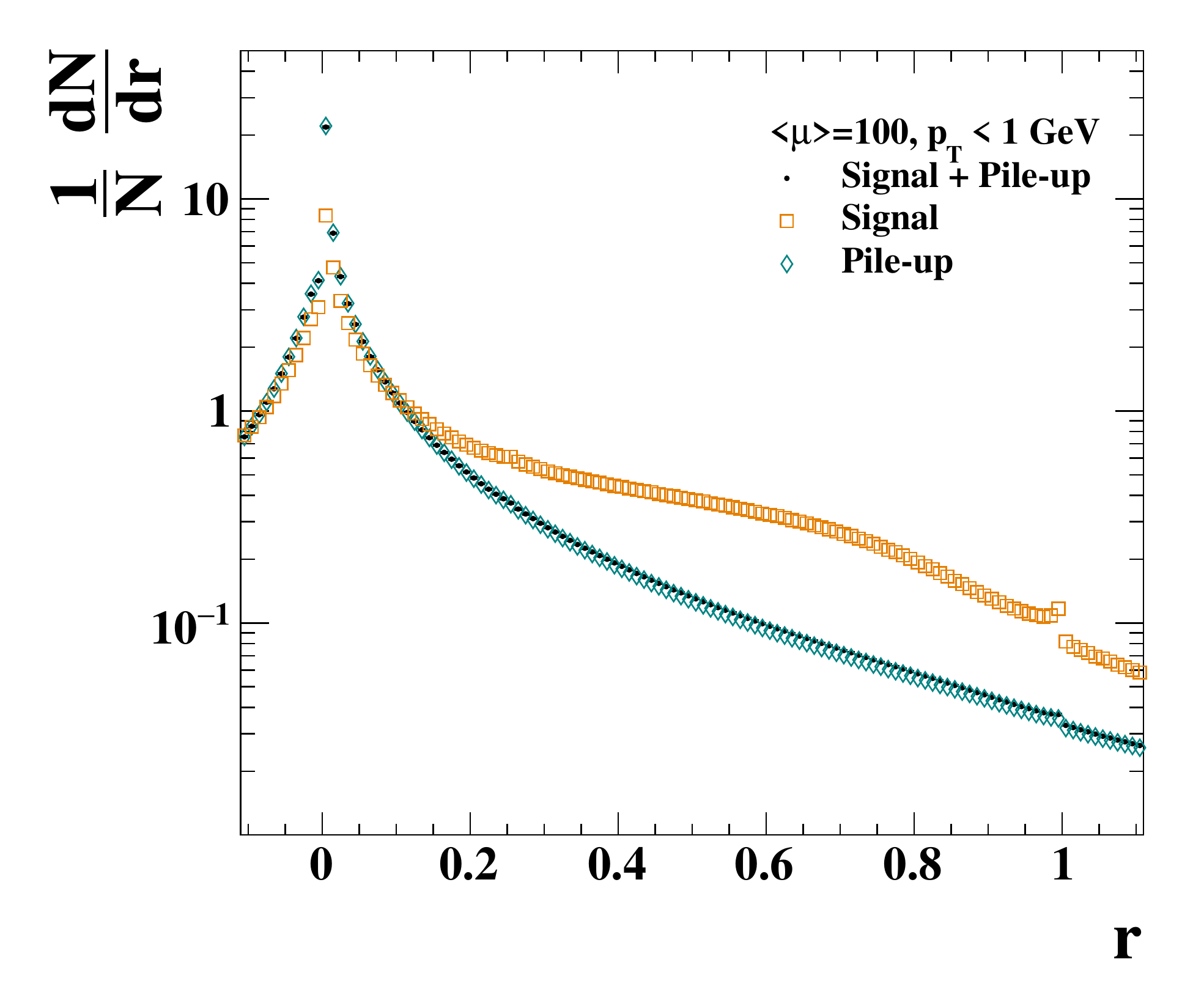}
    \vspace{-30pt}
    \caption{}
    \label{fig:rtracks_b}
  \end{subfigure}
  \caption[Distribution of r values] {Distribution of particle $r$ values in \Znunu\ events with \avnp~=~100 when using the dynamic threshold algorithm (Section~\ref{sec:dynamic}) and the Haar wavelet basis. (a) shows the distribution in which each particle enters with a weight equal to its \pT, and (b) shows the unweighted distribution for particles that satisfy $p_{T} < $~1~GeV.}
  \label{fig:rtracks}
\end{figure}

Cumulative distribution functions (CDFs) produced by integrating Figures~\ref{fig:rvalues}, \ref{fig:rByPT} and \ref{fig:rtracks} reveal the number of signal and pile-up particles, as well as the total \pT\ sum remaining as a function of the $r_{\mathrm{cut}}$ selection criterion that is applied. The signal and pile-up particle multiplicity as a function of $r_{\mathrm{cut}}$ is shown inclusively for all particles in Figure~\ref{fig:CDFMult}(a) and for particles that satisfy \pT~$<$~1~GeV, 1~$<$~\pT~$<$~4~GeV and \pT~$>$~4~GeV in Figure~\ref{fig:CDFMult}(b), (c) and (d), respectively. The fraction of signal particles in the event as a function of $r_{\mathrm{cut}}$ is shown below each distribution. The sheer number of low \pT\ pile-up particles means that the enhancement in the signal/pile-up ratio only becomes apparent for \pT~$>$~1~GeV.  However, when only those particles satisfying \pT~$>$~4~GeV are considered, as in Figure~\ref{fig:CDFMult}(\subref{fig:CDFMult_d}), the total number of signal particles is fairly constant regardless of the $r_{\mathrm{cut}}$ value used, while the number of pile-up particles drops very rapidly once $r_{\mathrm{cut}}\gtrapprox$~ 0.5.  This means that the total number of particles satisfying \pT~$>$~4~GeV  is dominated by the signal process for  $r_{\mathrm{cut}}\gtrapprox$~ 0.6 in this example where \avnp =~100.

When the CDF of the sum of particle \pT\ is used in Figure~\ref{fig:CDFPT} instead of particle multiplicity, the total \pT\ sum can be seen to be dominated by the signal contributions for $r_{\mathrm{cut}}$ values of around 0.7 and higher. The contribution from particles with \pT~$>$~4~GeV becomes signal-dominated at lower $r_{\mathrm{cut}}$ values, and the signal contribution from particles satisfying \pT~$>$~4~GeV is largely unaffected by the choice of $r_{\mathrm{cut}}$.  Thus the choice of $r_{\mathrm{cut}}$ value mainly affects the extent to which lower \pT\ particles are retained or rejected. The trade off between rejecting and retaining soft signal and pile-up particles is an inevitable feature of any pile-up mitigation method.

\begin{figure}[t!]
  \captionsetup[subfigure]{singlelinecheck=off}
  \begin{subfigure}{0.5\textwidth}
    \centering
    \includegraphics[width=1.0\textwidth]{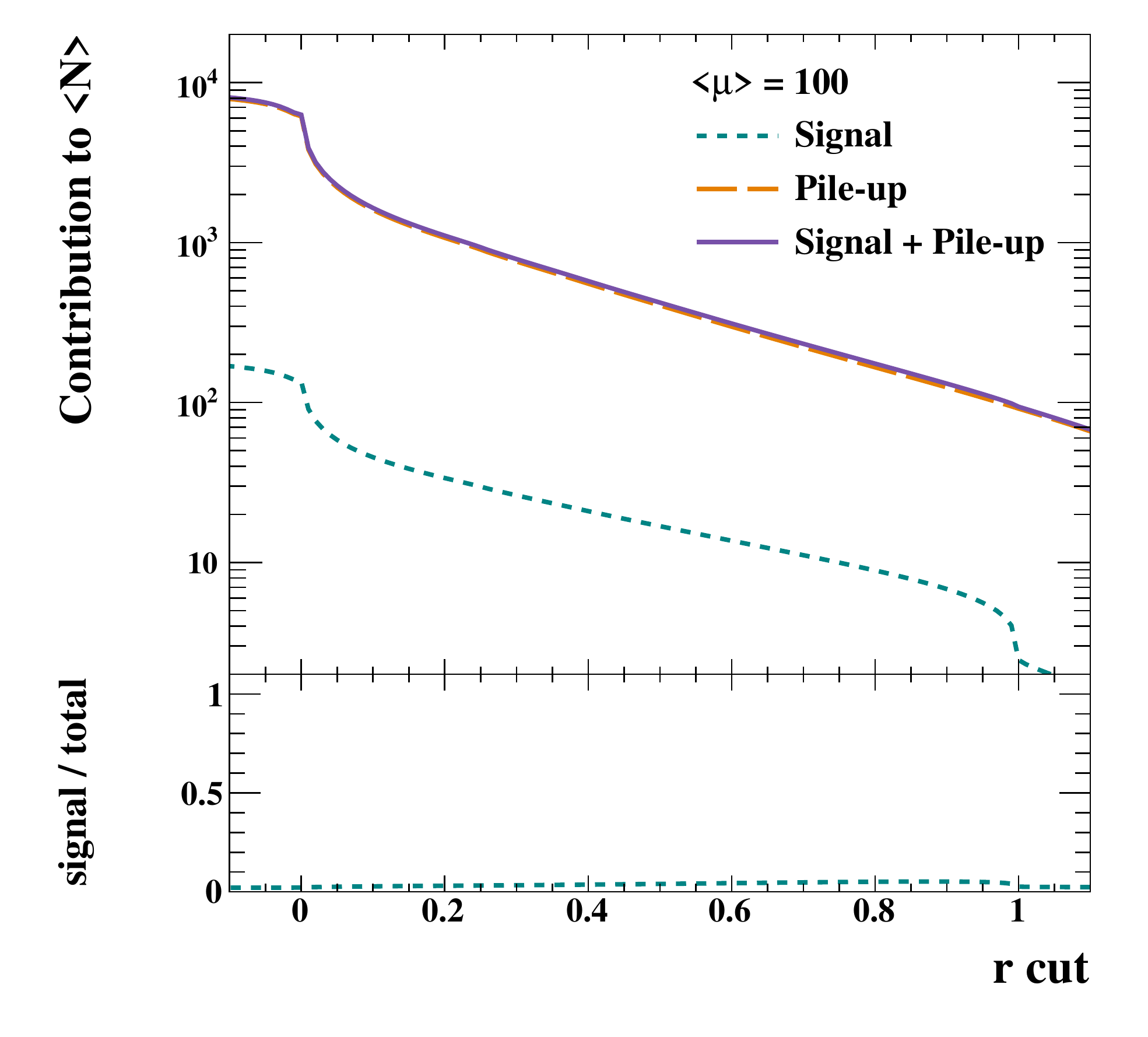}
    \vspace{-30pt}
    \caption{}
    \label{fig:CDFMult_a}
  \end{subfigure}
    \begin{subfigure}{0.5\textwidth}
    \centering
    \includegraphics[width=1.0\textwidth]{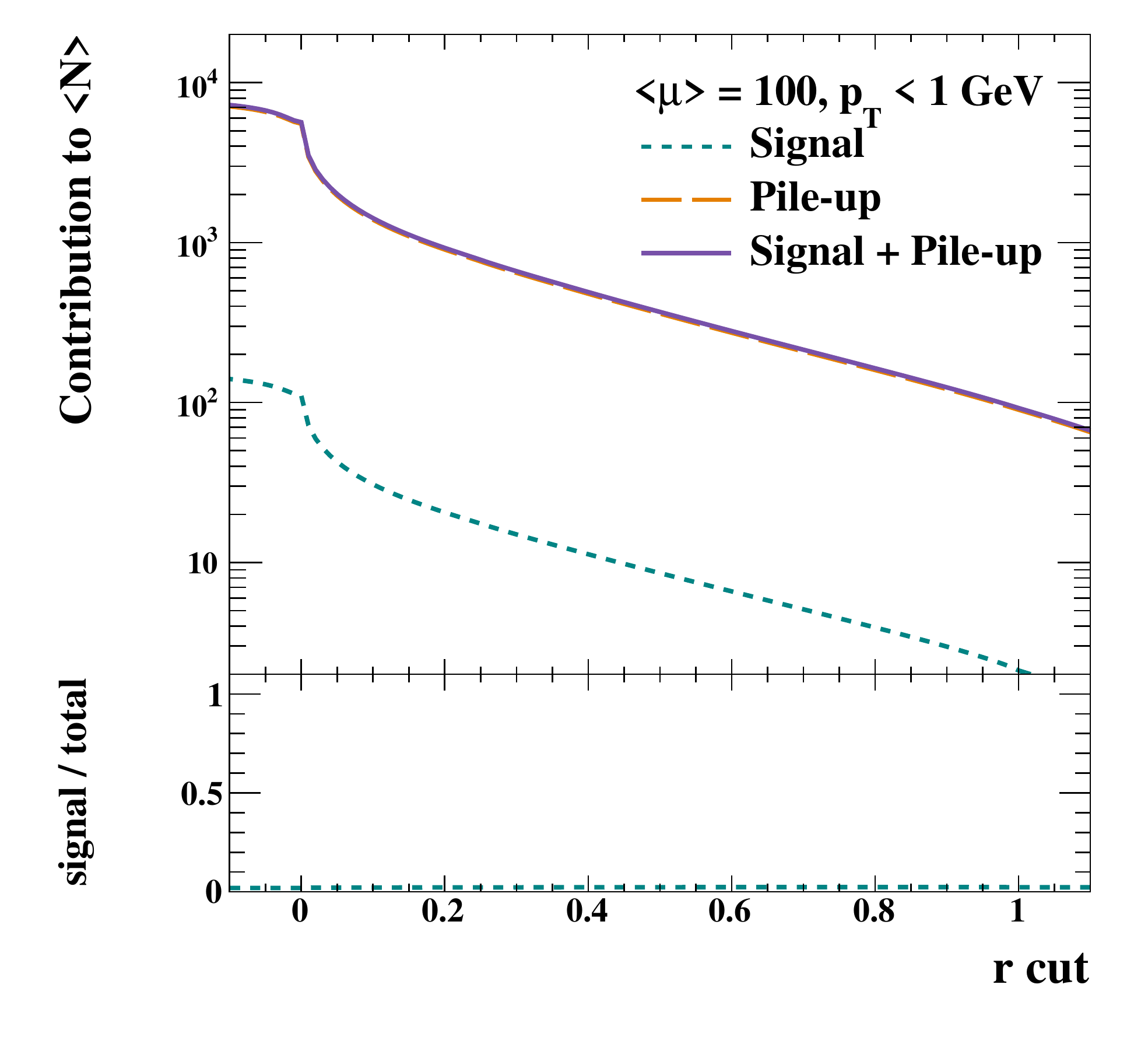}
    \vspace{-30pt}
    \caption{}
    \label{fig:CDFMult_b}
  \end{subfigure}\\
    \begin{subfigure}{0.5\textwidth}
    \centering
    \includegraphics[width=1.0\textwidth]{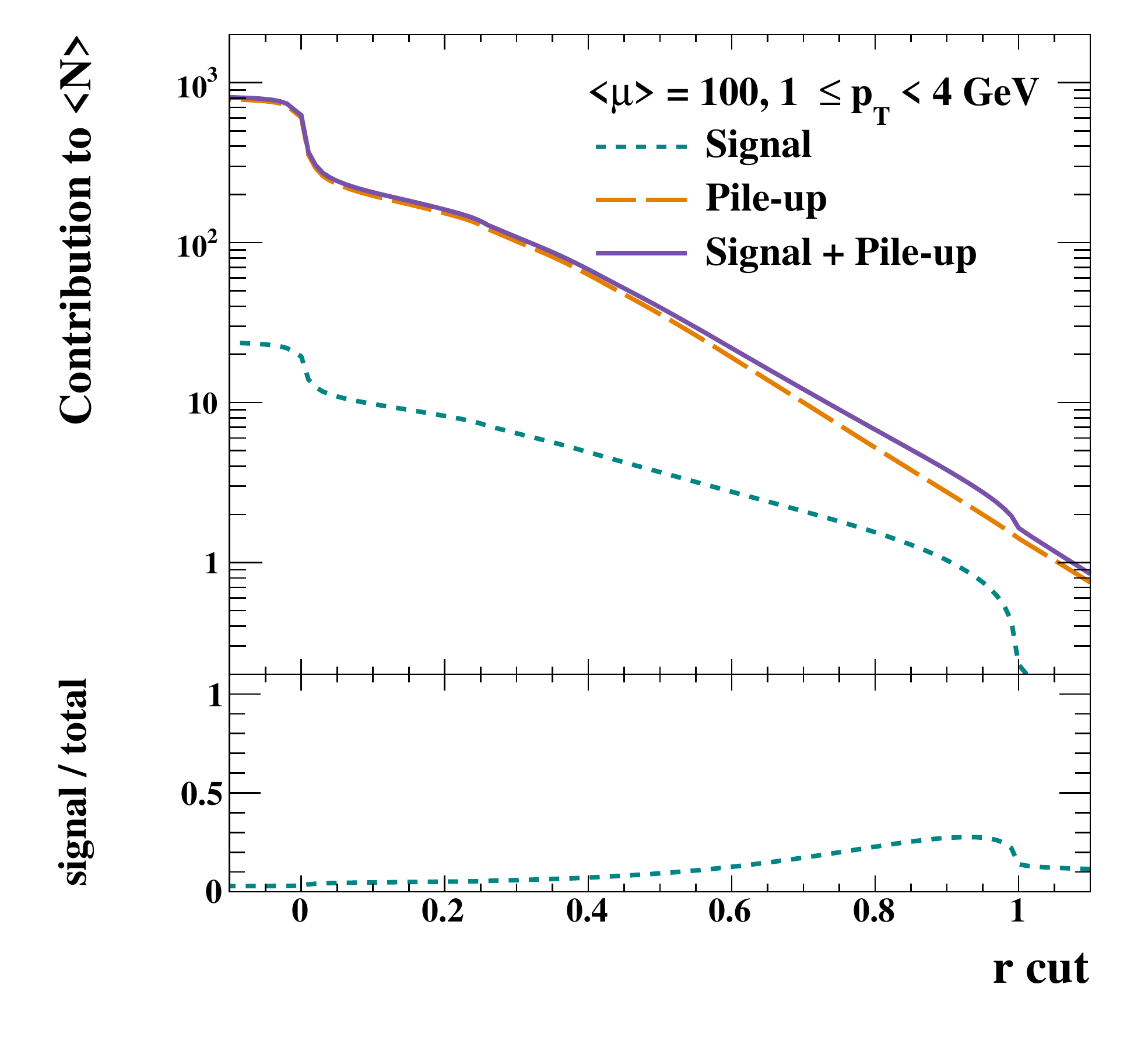}
    \vspace{-30pt}
    \caption{}
    \label{fig:CDFMult_c}
  \end{subfigure}
  \begin{subfigure}{0.5\textwidth}
    \centering
    \includegraphics[width=1.0\textwidth]{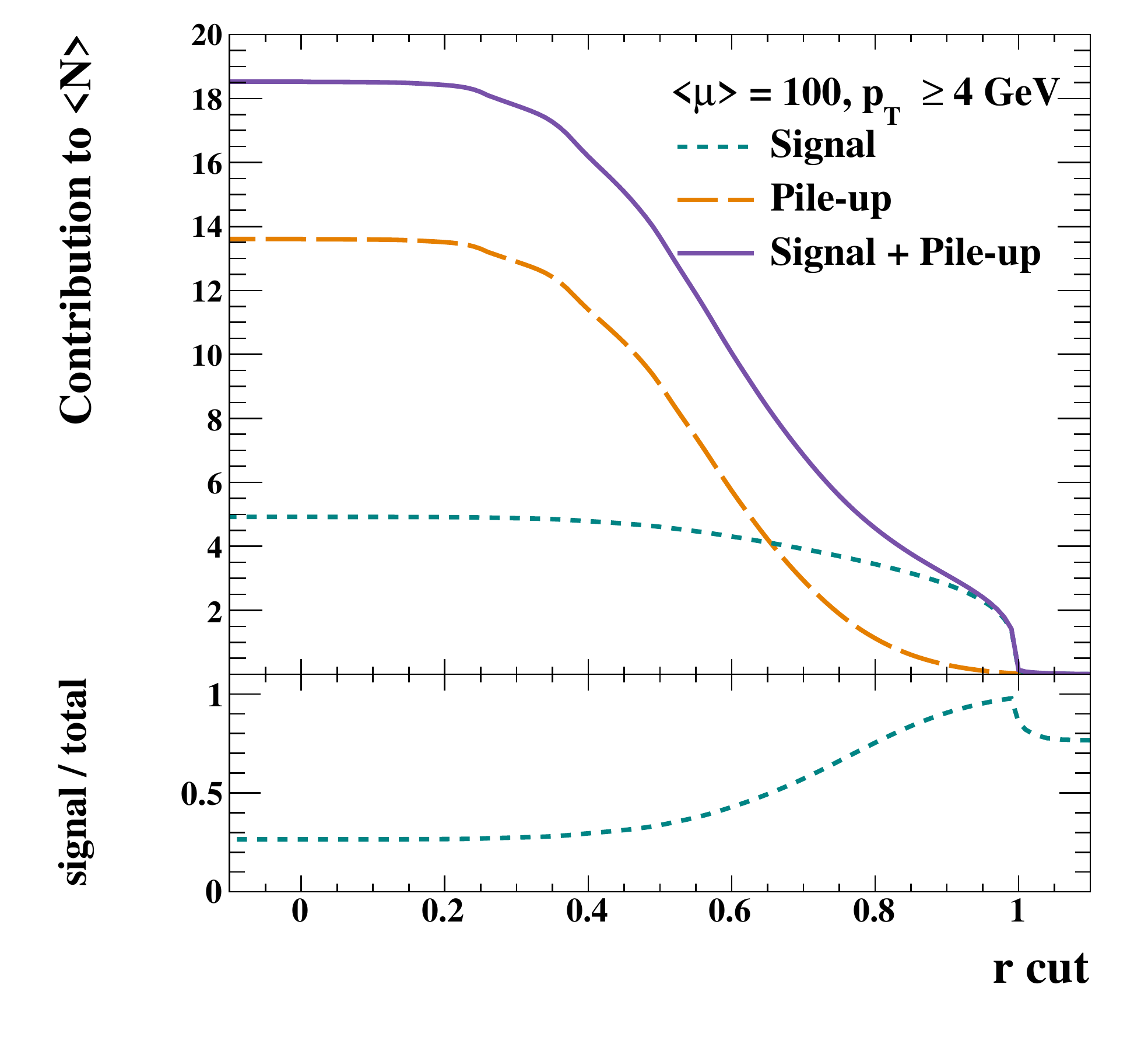}
    \vspace{-30pt}
    \caption{}
    \label{fig:CDFMult_d}
  \end{subfigure}
  \caption[$r_{\mathrm{cut}}$ dependence of average particle multiplicity] {The dependence of the average particle multiplicity, $\left<N\right>$, on $r_{\mathrm{cut}}$ in \Znunu\ events with \avnp~=~100 when using the flat-threshold algorithm (Section~\ref{sec:flat}) and the Haar wavelet basis. The multiplicities of the signal and pile-up particles are shown separately for all particles (a), and for those that satisfy \pT~$<$~1~GeV, 1~$<$~\pT~$<$~4~GeV, and \pT~$>$~4~GeV in (b), (c) and (d), respectively.}
  \label{fig:CDFMult}
\end{figure}

\begin{figure}[t!]
  \captionsetup[subfigure]{singlelinecheck=off}
  \begin{subfigure}{0.5\textwidth}
    \centering
    \includegraphics[width=1.0\textwidth]{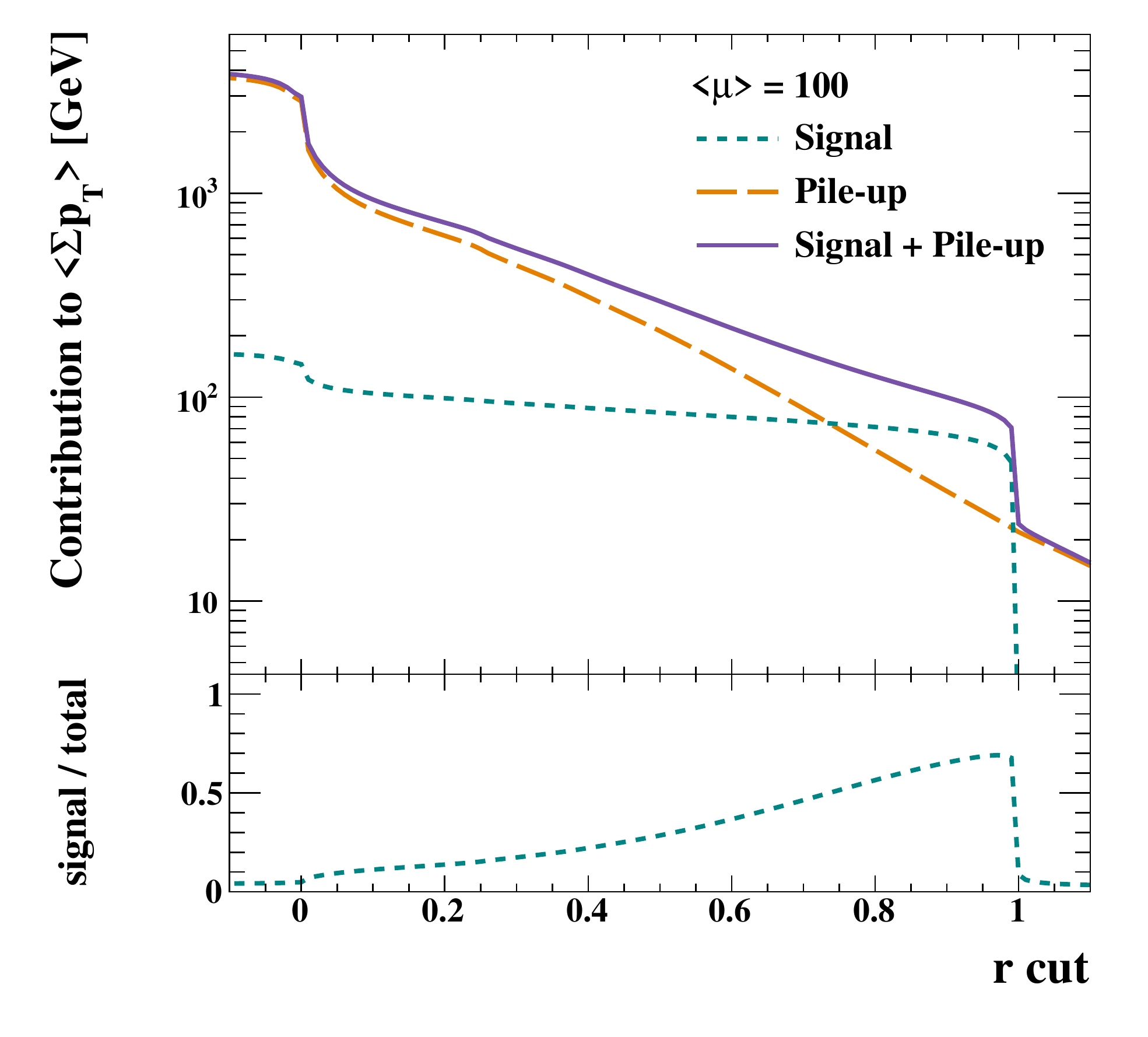}
    \vspace{-30pt}
    \caption{}
    \label{fig:CDFPT_a}
  \end{subfigure}
    \begin{subfigure}{0.5\textwidth}
    \centering
    \includegraphics[width=1.0\textwidth]{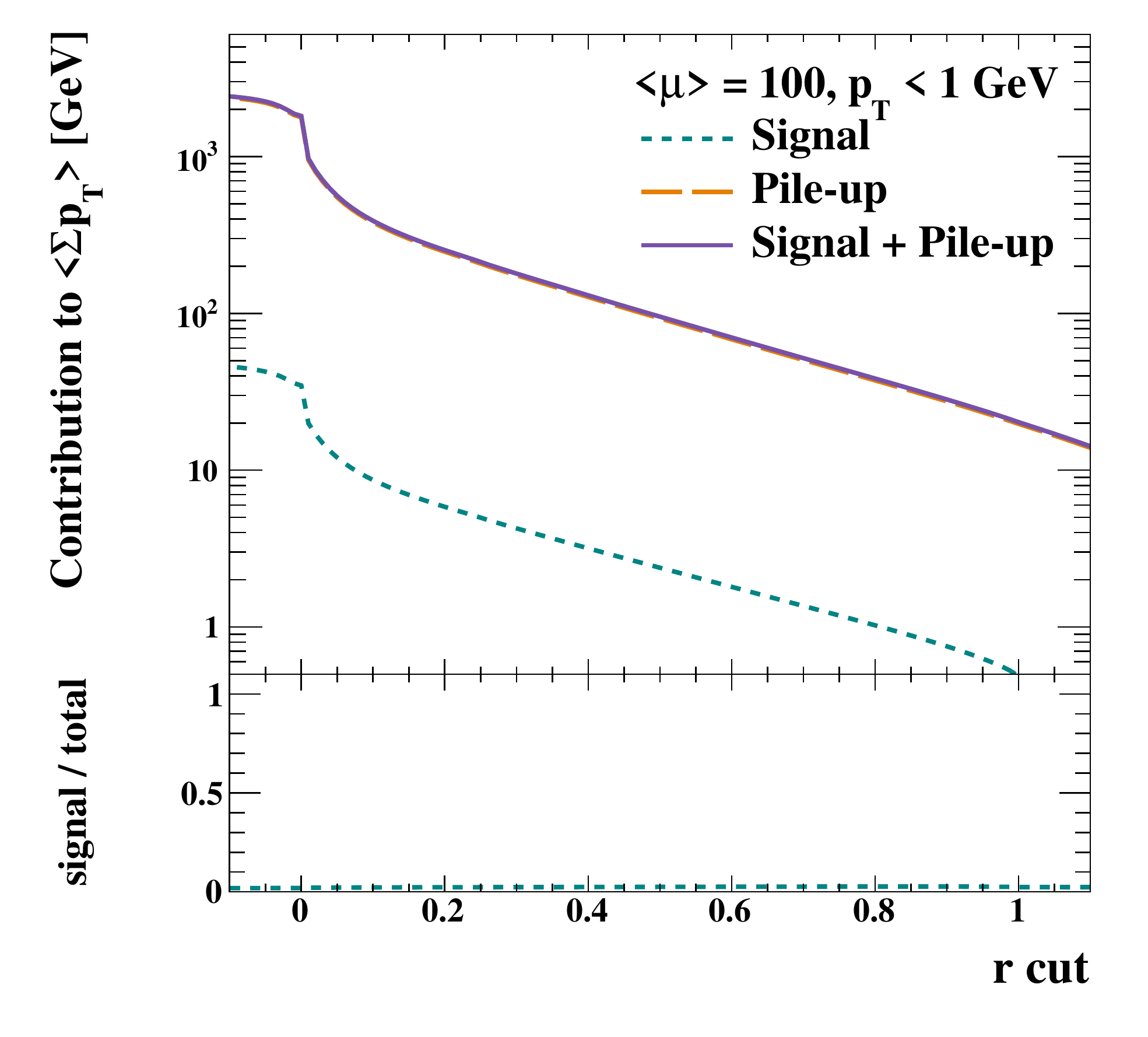}
    \vspace{-30pt}
    \caption{}
    \label{fig:CDFPT_b}
  \end{subfigure}\\
    \begin{subfigure}{0.5\textwidth}
    \centering
    \includegraphics[width=1.0\textwidth]{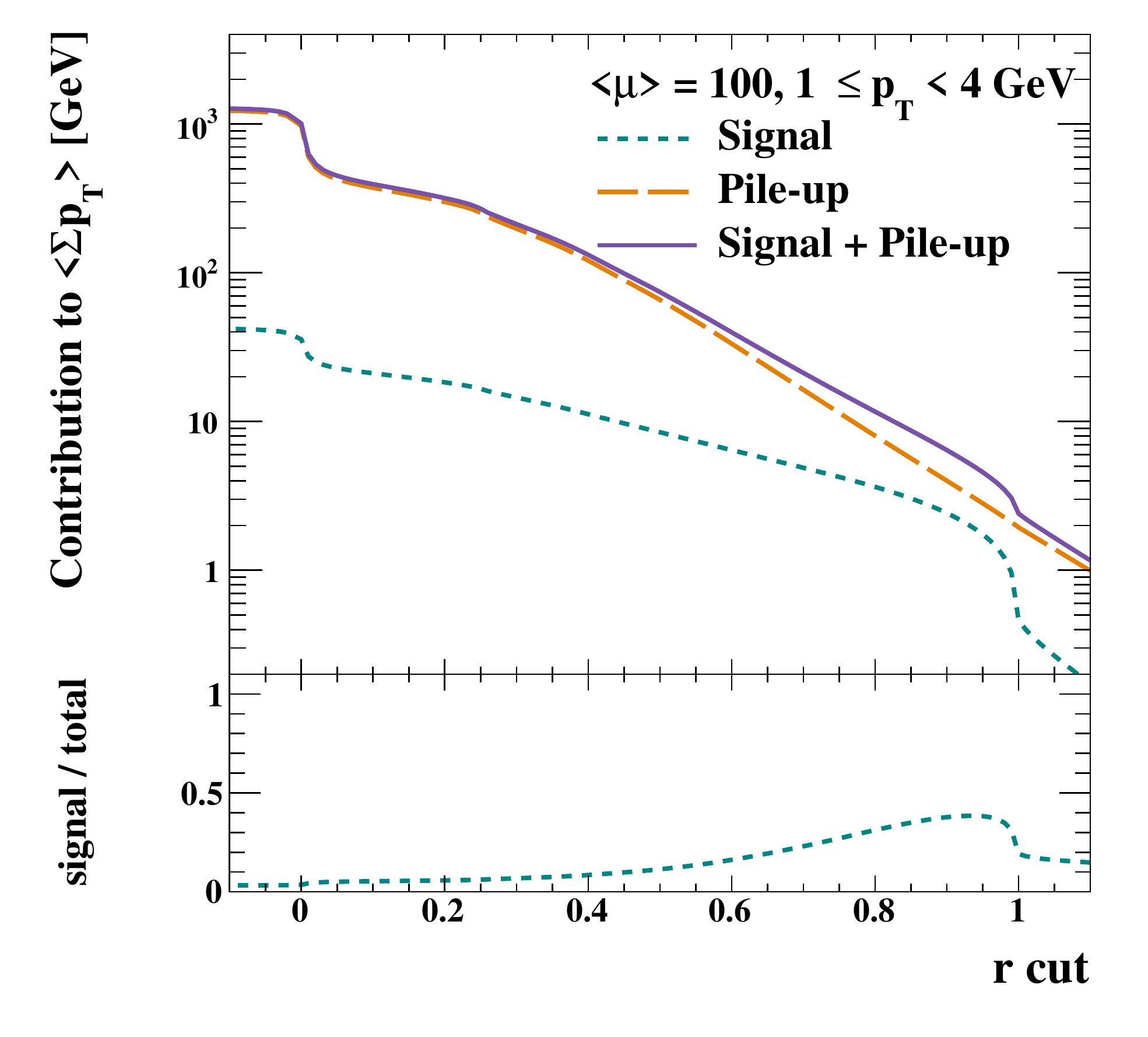}
    \vspace{-30pt}
    \caption{}
    \label{fig:CDFPT_c}
  \end{subfigure}
  \begin{subfigure}{0.5\textwidth}
    \centering
    \includegraphics[width=1.0\textwidth]{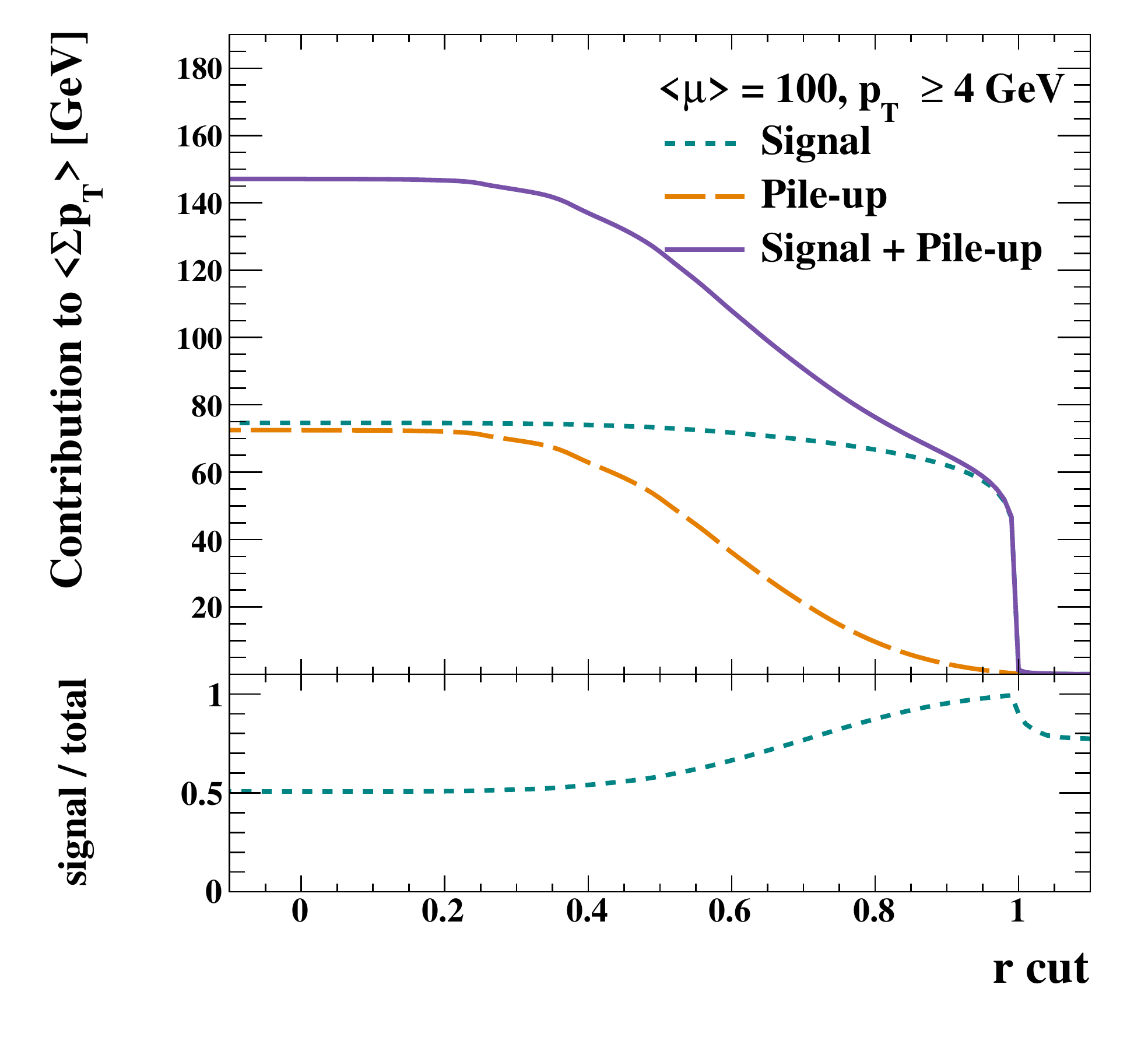}
    \vspace{-30pt}
    \caption{}
    \label{fig:CDFPT_d}
  \end{subfigure}
  \caption[$r_{\mathrm{cut}}$ dependence of average $\sum p_{T}$] {The dependence of the average particle $\sum p_{T}$ on $r_{\mathrm{cut}}$ in \Znunu\ events with \avnp~=~100 when using the flat-threshold algorithm (Section~\ref{sec:flat}) and the Haar wavelet basis. The contributions from signal and pile-up particles are shown separately for all particles (a), and for those that satisfy \pT~$<$~1~GeV, 1~$<$~\pT~$<$~4~GeV, and \pT~$>$~4~GeV in (b), (c) and (d), respectively.}
    \label{fig:CDFPT}
\end{figure}

Using the dynamic threshold algorithm shows an enhancement in the signal/pile-up ratio at low \pT, as revealed in the CDFs of Figures~\ref{fig:CDFMultTracks} and \ref{fig:CDFPTTracks}.   For particles with moderate \pTs\ between 1 and 4~GeV the event becomes signal dominated for $r_{\mathrm{cut}}$ values above around 0.6, an improvement compared with Figures~\ref{fig:CDFMult}(c) and \ref{fig:CDFPT}(c), which are not signal dominated at any $r_{\mathrm{cut}}$ value. For higher \pT\ particles with \pT~$>$~4~GeV, the pile-up rejection is even better, while almost all the signal is retained. 

\begin{figure}[t!]
  \captionsetup[subfigure]{singlelinecheck=off}
  \begin{subfigure}{0.5\textwidth}
    \centering
    \includegraphics[width=1.0\textwidth]{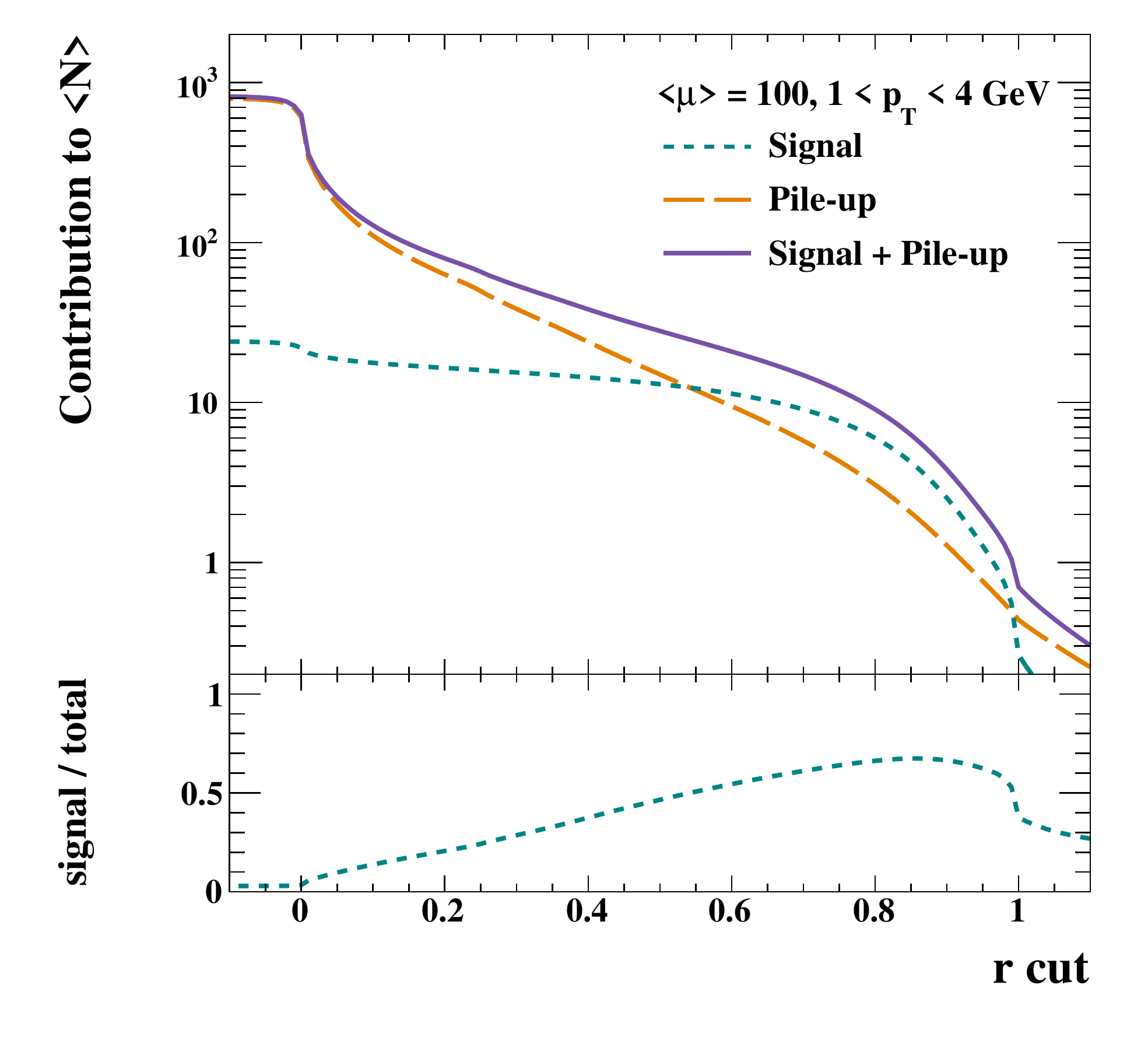}
    \vspace{-30pt}
    \caption{}
    \label{fig:CDFPTTracks_a}
  \end{subfigure}
  \begin{subfigure}{0.5\textwidth}
    \centering
    \includegraphics[width=1.0\textwidth]{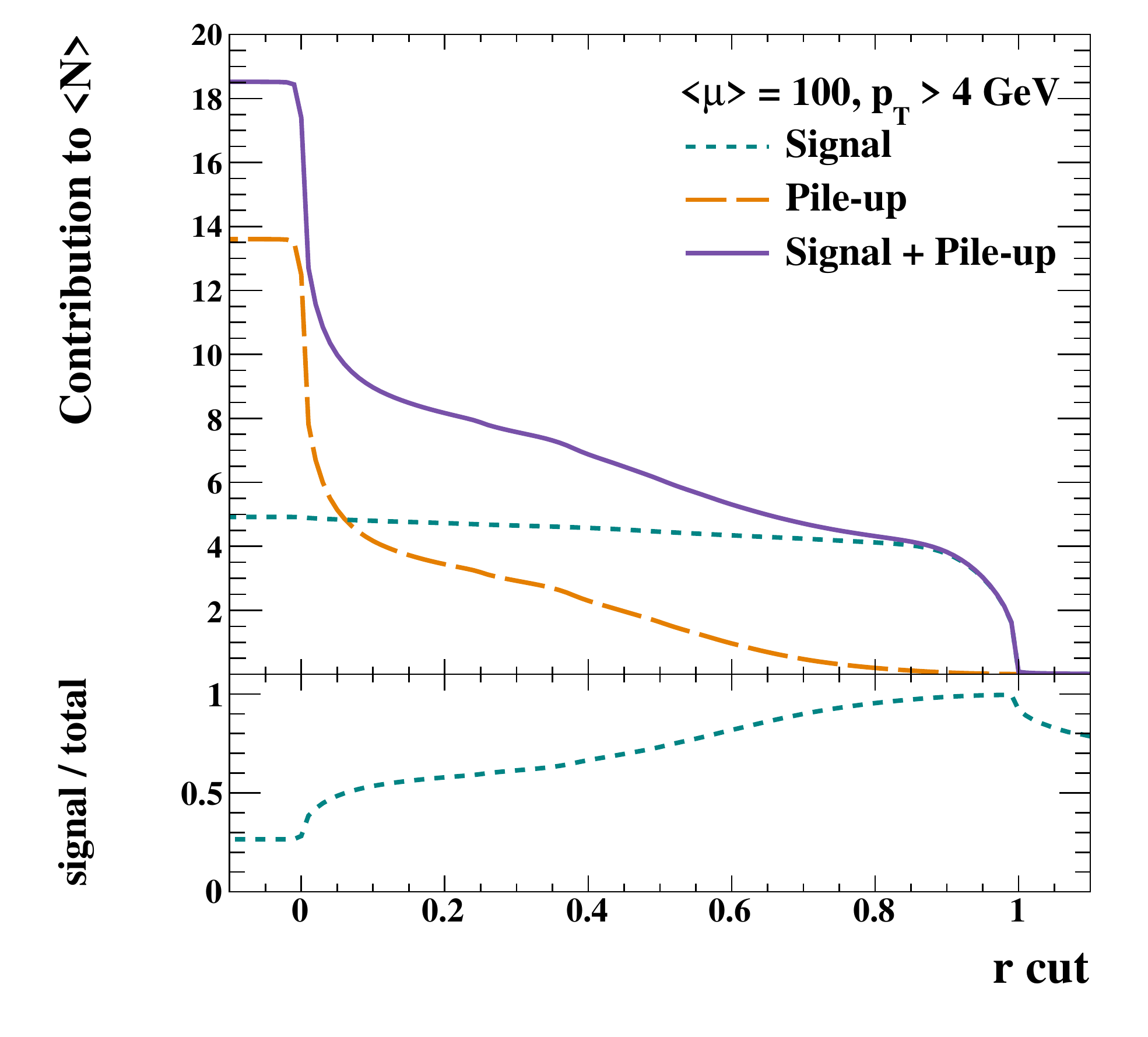}
    \vspace{-30pt}
    \caption{}
    \label{fig:CDFPTTracks_b}
  \end{subfigure}
  \caption[Dependence of $\sum p_{T}$ on $r_{\mathrm{cut}}$] {The dependence of the average particle multiplicity, $\left<N\right>$, on $r_{\mathrm{cut}}$ in $Z\rightarrow\nu\bar{\nu}$ events with \avnp~=~100 when using the dynamic threshold algorithm (Section \ref{sec:dynamic})  and the Haar wavelet basis.  The multiplicities of the signal and pile-up particles are shown separately for particles that satisfy 1~$<$~\pT~$<$~4~GeV (a), and \pT~$>$~4~GeV (b).}
  \label{fig:CDFMultTracks}
\end{figure}

\begin{figure}[t!]
  \captionsetup[subfigure]{singlelinecheck=off}
  \begin{subfigure}{0.5\textwidth}
    \centering
    \includegraphics[width=1.0\textwidth]{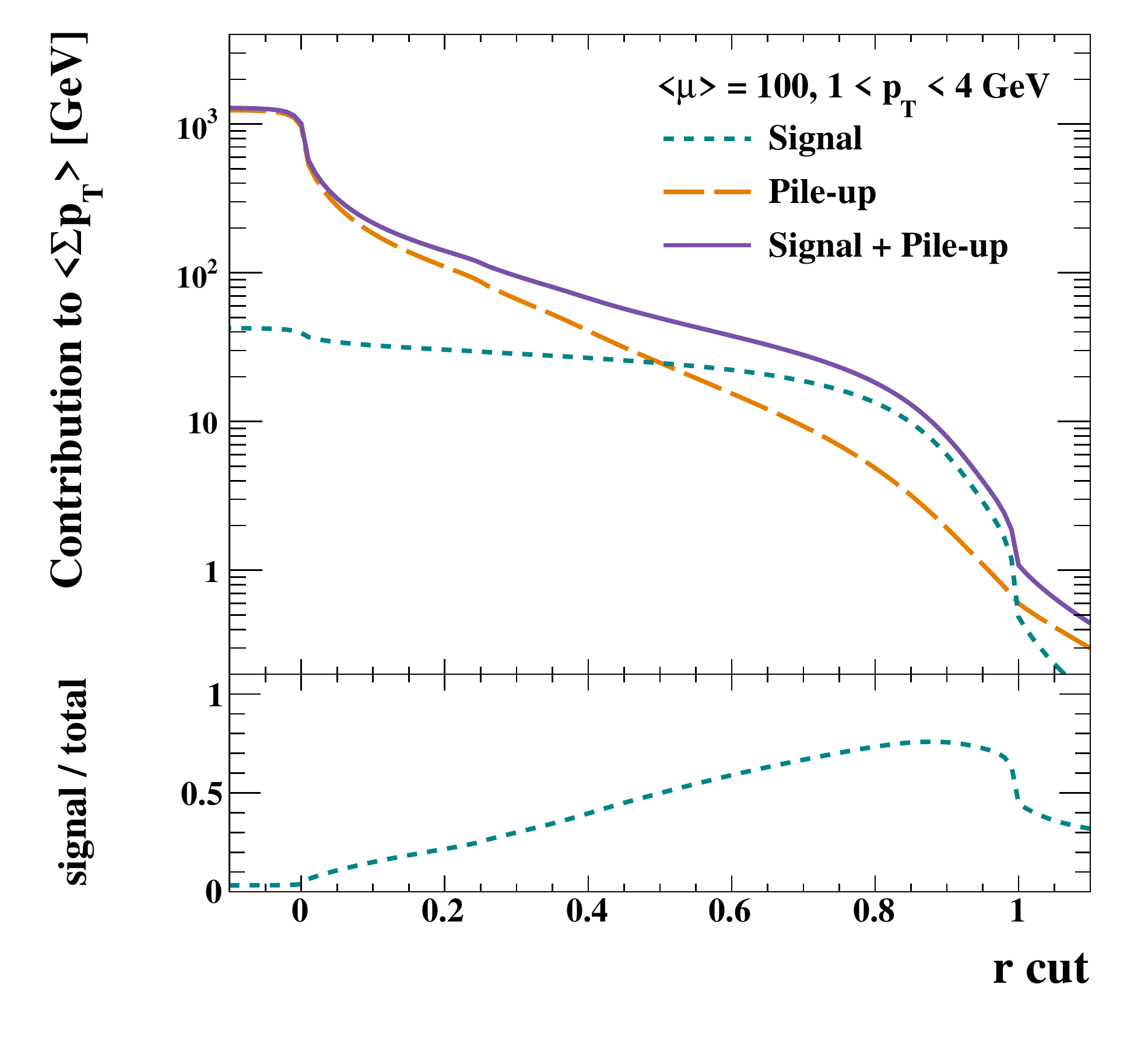}
    \vspace{-30pt}
    \caption{}
    \label{fig:CDFPTTracks_a}
  \end{subfigure}
  \begin{subfigure}{0.5\textwidth}
    \centering
    \includegraphics[width=1.0\textwidth]{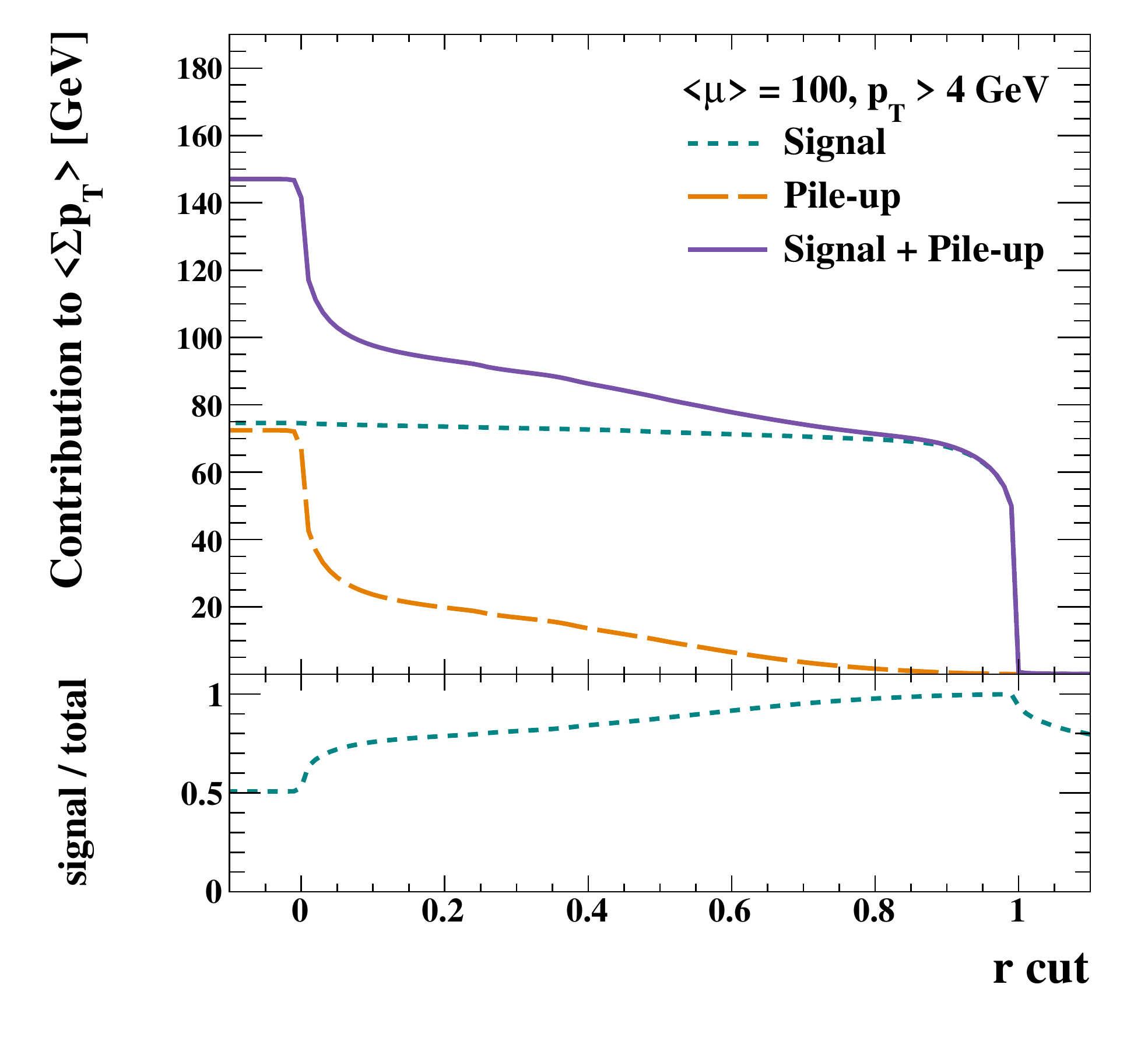}
    \vspace{-30pt}
    \caption{}
    \label{fig:CDFPTTracks_b}
  \end{subfigure}
  \caption[Dependence of $\sum p_{T}$ on $r_{\mathrm{cut}}$] {The dependence of the average particle $\sum p_{T}$ on $r_{\mathrm{cut}}$ in \Znunu\ events with \avnp~=~100 when using the dynamic threshold algorithm (Section~\ref{sec:dynamic}) and the Haar wavelet basis. The contributions from signal and pile-up particles are shown separately for particles that satisfy 1~$<$~\pT~$<$~4~GeV (a), and \pT~$>$~4~GeV (b).}
  \label{fig:CDFPTTracks}
\end{figure}

Receiver Operator Characteristic (ROC) curves show the contours of expected signal retention versus pile-up rejection for both the flat-threshold algorithm in Figure~\ref{fig:ROC} and the dynamic threshold algorithm in Figure~\ref{fig:ROCTracks} using the Haar wavelet basis. The best performing regions are in the upper left of the plots where the signal retention is as close to one as possible and the pile-up retention is close to zero. ROC curves are produced for \avnp\ values of 50, 100 and 300, and the left hand panels show the performance for particle multiplicity, while the right hand panels show the performance for the particle \pT\ sum. The curves are separated into the three different \pT\ regions of \pT~$<$~1~GeV, 1~$<$~\pT~$<$~4~GeV and 4~GeV~$<$~\pT, and as before, the performance is higher for higher \pT\ particles, and is not strongly dependent on \avnp. The addition of charged particle tracking information improves the ROC curves at all \avnp\ values and \pT\ regions.
\newpage
\begin{figure}[t!]
  \captionsetup[subfigure]{singlelinecheck=off}
  \begin{subfigure}{1.0\textwidth}
    \centering
    \includegraphics[width=0.95\textwidth]{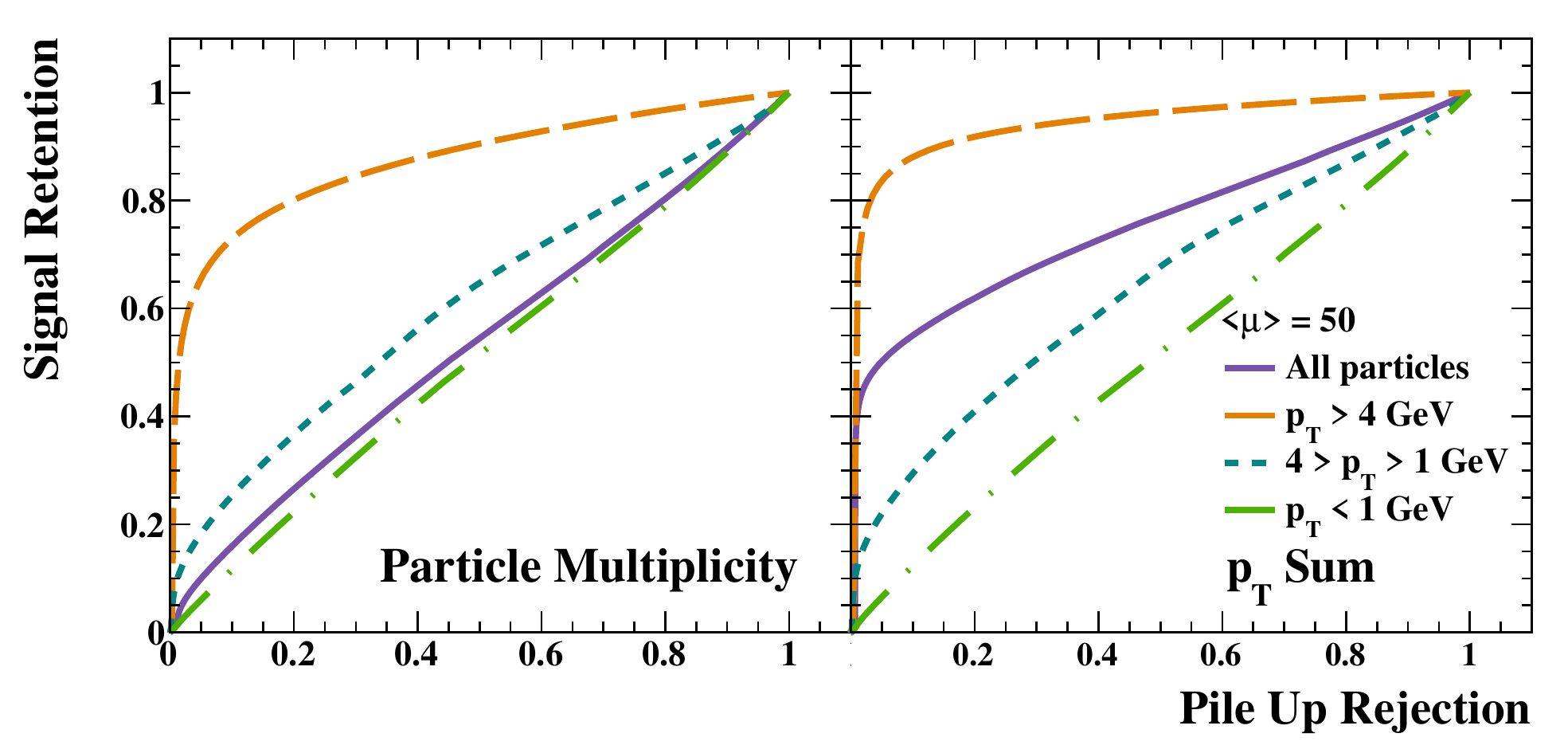}
    \vspace{-30pt}
    \caption{}
    \label{fig:ROC_a}
    \vspace{18pt}
  \end{subfigure}
  \begin{subfigure}{1.0\textwidth}
    \centering
    \includegraphics[width=0.95\textwidth]{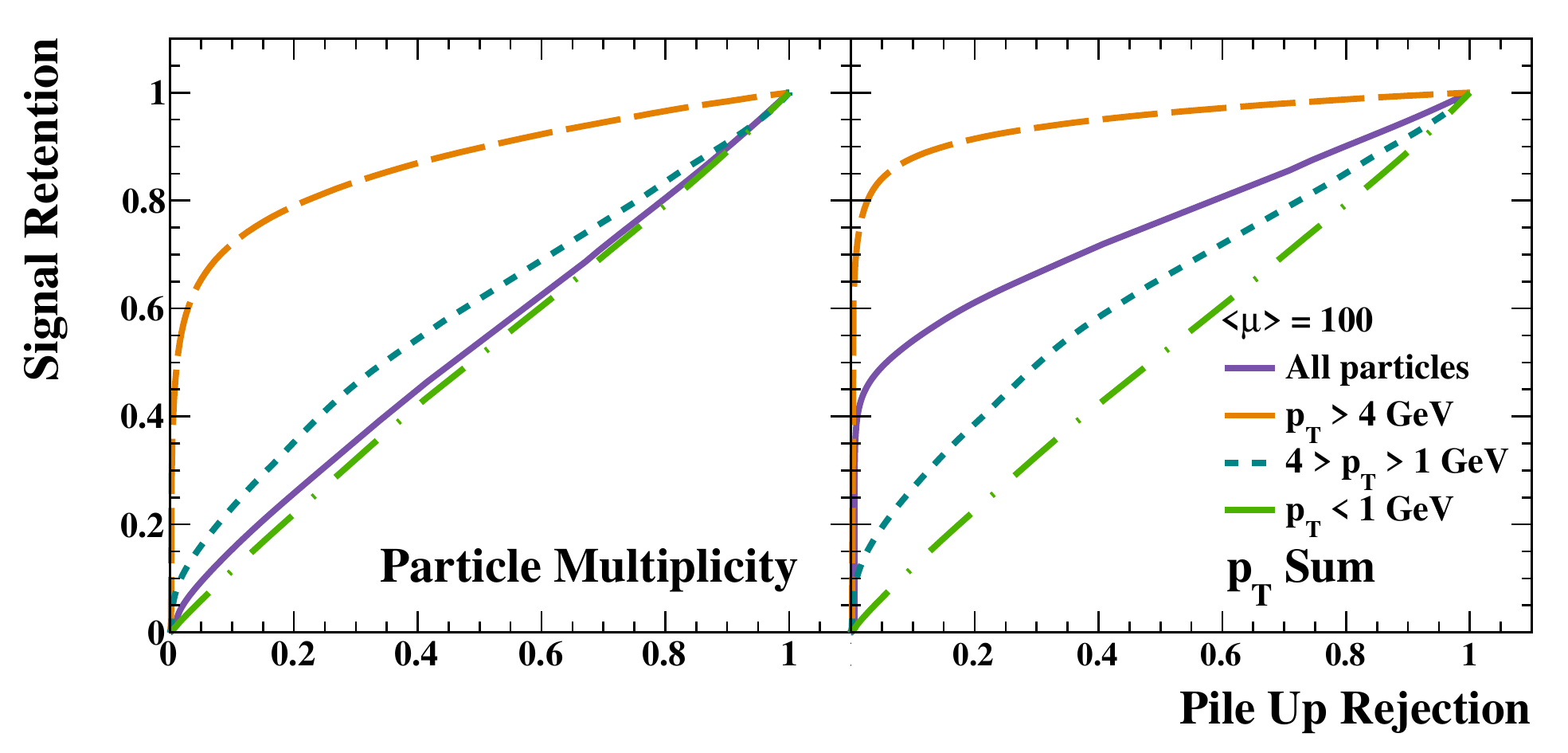}
    \vspace{-30pt}
    \caption{}
    \label{fig:ROC_b}
    \vspace{18pt}
  \end{subfigure}
  \begin{subfigure}{1.0\textwidth}
    \centering
    \includegraphics[width=0.95\textwidth]{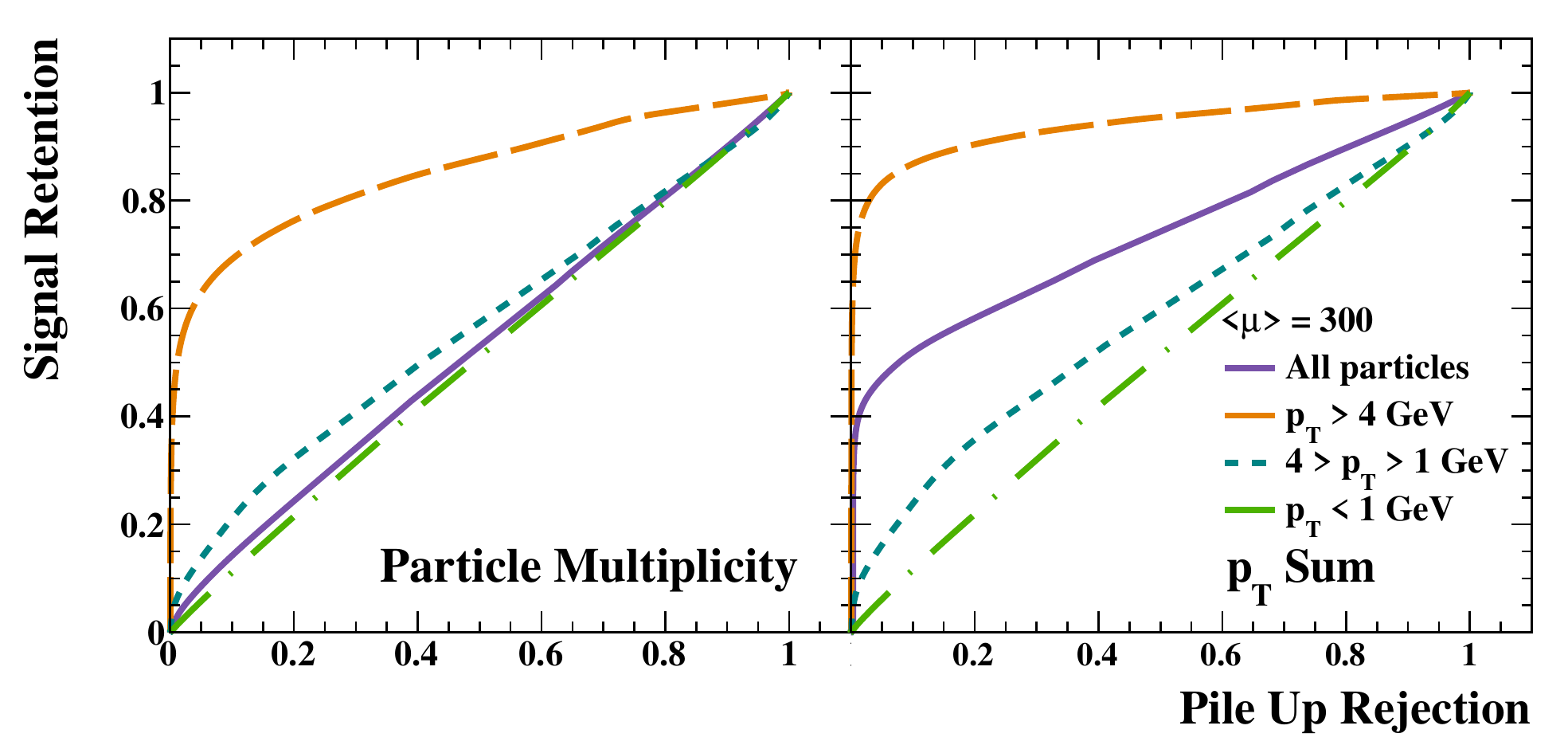}
    \vspace{-30pt}
    \caption{}
    \label{fig:ROC_c}
    \vspace{18pt}
  \end{subfigure}
  \caption[ROC curves] {ROC curves when using the flat-threshold algorithm (Section~\ref{sec:flat}) at \avnp=50 (top), 100 (middle) and 300 (bottom) for particle multiplicity (left) and the sum of particle \pT\ (right). Curves are shown for all particles, as well as three different \pT\ regions.}
  \label{fig:ROC}
\end{figure}
\begin{figure}[t!]
  \captionsetup[subfigure]{singlelinecheck=off}
  \begin{subfigure}{1.0\textwidth}
    \centering
    \includegraphics[width=0.95\textwidth]{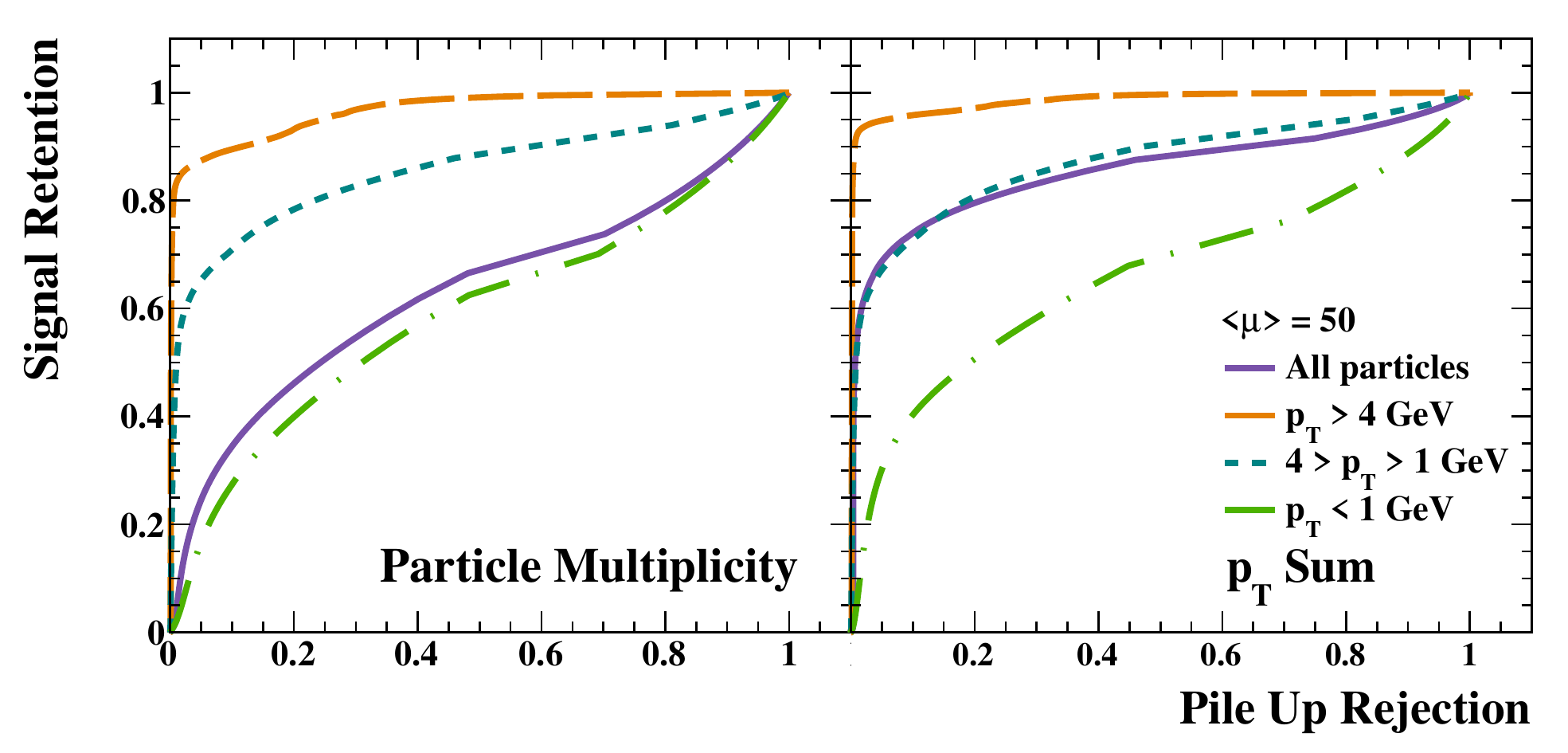}
    \vspace{-30pt}
    \caption{}
    \label{fig:ROCTracks_a}
    \vspace{18pt}
  \end{subfigure}
  \begin{subfigure}{1.0\textwidth}
    \centering
    \includegraphics[width=0.95\textwidth]{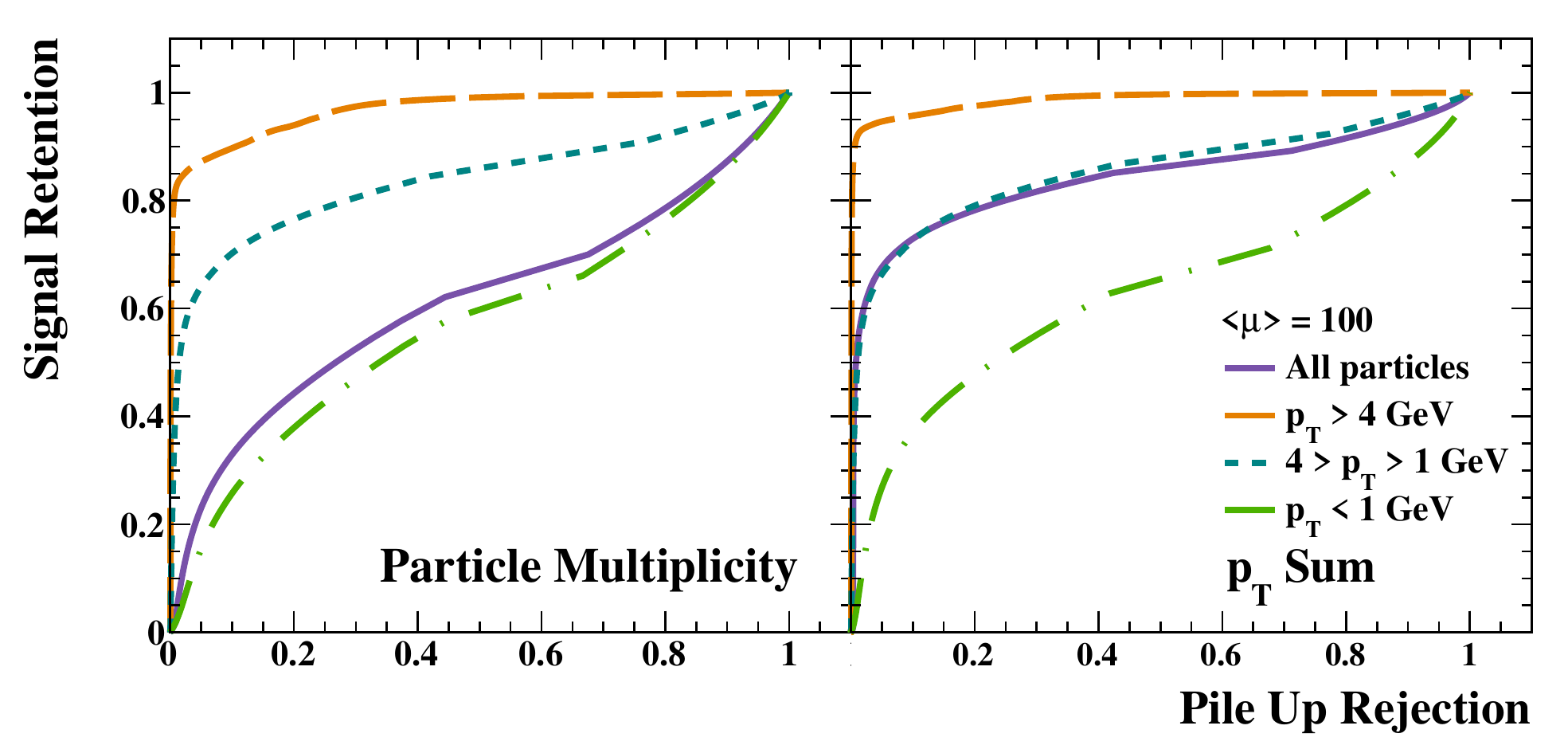}
    \vspace{-30pt}
    \caption{}
    \label{fig:ROCTracks_b}
    \vspace{18pt}
  \end{subfigure}
  \begin{subfigure}{1.0\textwidth}
    \centering
    \includegraphics[width=0.95\textwidth]{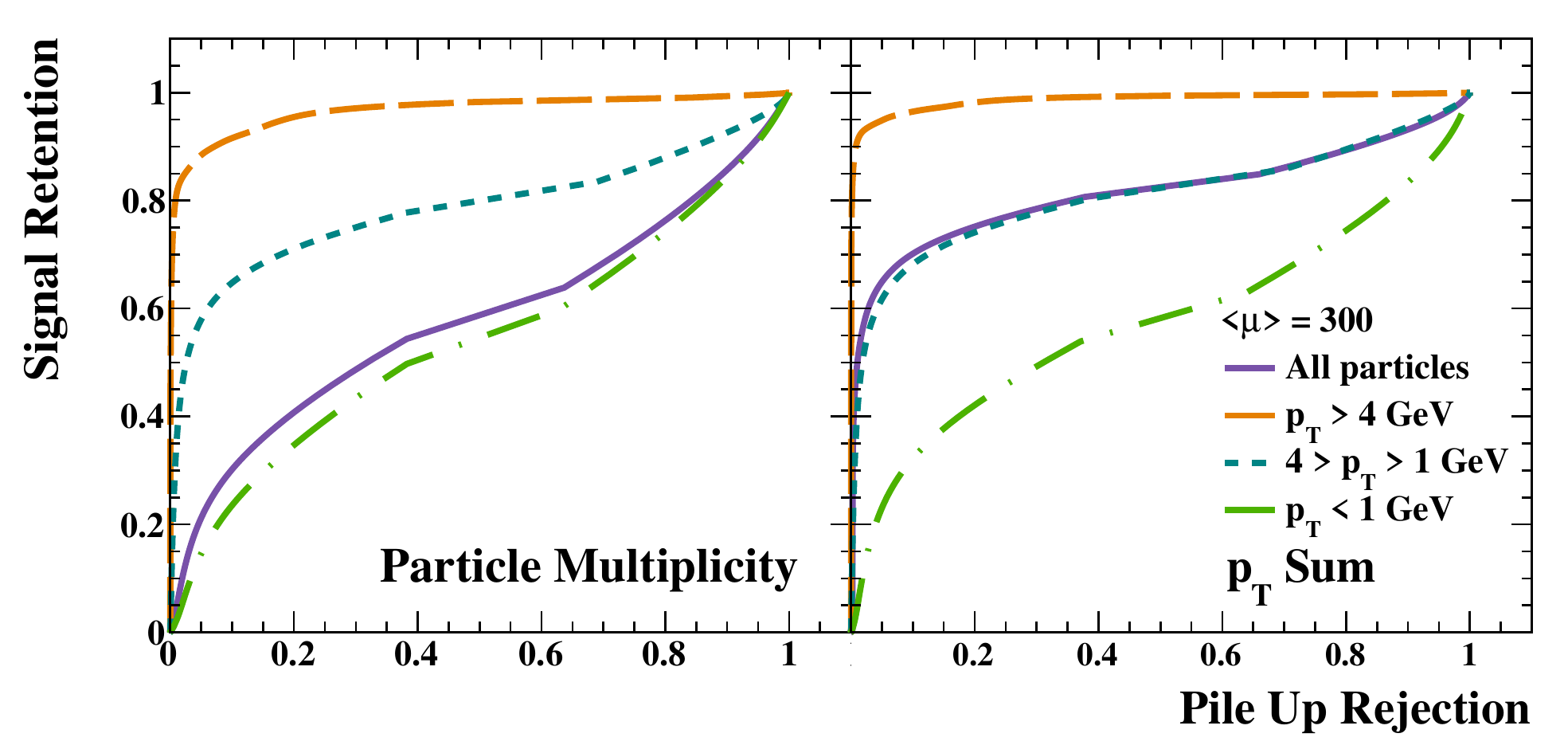}
    \vspace{-30pt}
    \caption{}
    \label{fig:ROCTracks_c}
    \vspace{18pt}
  \end{subfigure}
  \caption[ROC curves] {ROC curves when using the dynamic threshold algorithm (Section~\ref{sec:dynamic}) at \avnp=50 (top), 100 (middle) and 300 (bottom) for particle multiplicity (left) and the sum of particle \pT\ (right). Curves are shown for all particles, as well as three different \pT\ regions}
  \label{fig:ROCTracks}
\end{figure}

\section{Conclusion}

We have introduced a new method for pile-up mitigation based on wavelet decomposition of individual beam-crossing events.  The method is valid for $\mu$ values greater than approximately 20, and scales to very high $\mu$, as is appropriate for the high luminosity upgrade of the LHC and beyond.  The method models pile-up at high-$\mu$ as a form of white noise, whose fluctuations are Poissonian when viewed at all resolvable angular scales.

As with all pile-up mitigation schemes, the separation between signal and pile-up is worse for low-\pT particles.  Indeed, in the limit of a zero \pT particle, no method would be capable of distinguishing signal and background production.  However, for particles with moderate to high \pT, the method provides good separation between signal and pile-up, even in the challenging case of $Z\rightarrow\nu\bar{\nu}$ production used here.

The method can be further improved by adding information from charged particle tracks to help identify signal regions in the wavelet domain, which allows a tighter threshold to be used in order to reject more pile-up without removing signal particles.

The performance on the classification of individual particles is presented here, and shows that the method is capable of separating signal particles from pile-up even at quite low-\pT and high-$\mu$.  Further studies will evaluate the performance of the method on higher level observables, which will permit a comparison to other existing pile-up mitigation approaches.

\section*{Acknowledgments}
Thanks to Troels Petersen for useful discussions on the application of the method.  This work was funded by the Danish National Research Foundation.
\bibliography{references}

\end{document}